\documentclass{aa}
\usepackage{upgreek}
\usepackage{graphicx}
\usepackage[varg]{txfonts}
\usepackage{natbib}
\usepackage{float}
\usepackage{soul}
\newlength{\linwx}
\usepackage{color}

\begin{document}

\title{Diversity of disc viscosities can explain the period ratios of resonant and non-resonant systems of hot super-Earths and mini-Neptunes}

\author{
Bertram Bitsch \inst{1,2} and Andre Izidoro \inst{3}
}
\offprints{B. Bitsch,\\ \email{bitsch@mpia.de}}
\institute{
Department of Physics, University College Cork, Cork, Ireland 
\and
Max-Planck-Institut f\"ur Astronomie, K\"onigstuhl 17, 69117 Heidelberg, Germany
\and
Department of Earth, Environmental and Planetary Sciences, MS 126, Rice
 University, Houston, TX 77005, USA
}
\abstract{Migration is a key ingredient in the formation of close-in super-Earth and mini-Neptune systems. The migration rate sets the resonances  in which planets can be trapped, where slower migration rates result in wider resonance configurations compared to higher migration rates. We investigate the influence of different migration rates ---set by disc viscosity--- on the structure of multi-planet systems via N-body simulations, where planets grow via pebble accretion. Planets in low-viscosity environments migrate slower due to partial gap opening compared to planets forming in high-viscosity environments. Consequently, systems formed in low-viscosity environments tend to have planets trapped in wider resonant configurations (typically 4:3, 3:2, and 2:1 configurations). Simulations of high-viscosity discs mostly produce planetary systems in 7:6, 5:4, and 4:3 resonances. After the gas disc dissipates, the damping forces of eccentricity and inclination cease to exist and the systems can undergo instabilities on timescales of a few tens of  millions of years, rearranging their configurations and breaking the resonance chains. We show that low-viscosity discs naturally account for the configurations of resonant chains, such as Trappist-1, TOI-178, and Kepler-223, unlike high-viscosity simulations, which produce chains that are more compact. Following dispersal of the gas disc, about 95\% of our low-viscosity resonant chains became unstable, experiencing a phase of giant impacts. Dynamical instabilities in our low-viscosity simulations are more violent than those of high-viscosity simulations due to the effects of leftover external perturbers (P$>$200 days). About 50\% of our final systems end with no planets within 200 days, while all our systems harbour remaining outer planets. We speculate that this process could be qualitatively consistent with the lack of inner planets in a large fraction of the Sun-like stars. Systems produced in low-viscosity simulations alone do not match the overall period ratio distribution of observations, but give a better match to the period distributions of chains, which may suggest that systems of super-Earths and mini-Neptunes form in natal discs with a diversity of viscosities.
}
\authorrunning{Bitsch and Izidoro}\titlerunning{Diversity of viscosities}\maketitle

\section{Introduction}
\label{sec:Introduction}

In contrast to the Solar System, a large fraction of exoplanetary systems harbour super-Earth/mini-Neptune planets in orbits interior to that of Mercury (e.g. \citealt{2011arXiv1109.2497M, 2013ApJ...766...81F, 2018AJ....156...24M}). These systems of close-in planets normally consist of several planets that can be in or close to mean-motion resonances (MMRs) (e.g. \citealt{2014ApJ...790..146F, 2019A&A...625A...7P}). Furthermore, it seems that the majority of these systems behave like `peas in a pod', where the planets have roughly equal sizes and there is an equal distribution of period ratios within each system \citep{2017ApJ...849L..33M, 2018AJ....155...48W, 2021ApJ...920L..34M}. However, their formation pathway is still under debate.

While it is clear that the formation of systems in resonance requires planetary migration (e.g. \citealt{2015A&A...578A..36O, 2017MNRAS.470.1750I, 2019arXiv190208772I, 2021A&A...656A.115H}), it is unclear whether the planets originate from beyond the water-ice line or formed in the inner regions of the disc. This is independent of whether or not the planetary systems are in resonance. For selected systems, the chemical composition of the planets could be used to decipher the location of their formation (e.g. \citealt{2019ApJ...887L..14B, 2024AJ....167..167S}), as the migration behaviour sets the composition of these planets (e.g. \citealt{2019A&A...624A.109B, 2019A&A...627A.149S, 2021A&A...649L...5B, 2022ApJ...939L..19I, 2024A&A...681A..52C}). However, even with JWST operating and the future ARIEL mission, we are still far away from a bulk characterization of planets in these systems. In addition, a spectroscopic analysis of the planetary atmospheres requires that the planets be transiting, and that they orbit stars close enough to allow a sufficient signal-to-noise ratio. However, for most of the \textit{Kepler} sample, this remains a challenge due to the large distance of the \textit{Kepler} field from Earth, leaving it unclear as to whether the majority of the close-in planets formed interior or exterior to the water ice line from a detailed compositional point of view.

On the other hand, the analysis of the \textit{Kepler} sample revealed a valley in the radius distribution of these close-in planets \citep{2017AJ....154..109F, 2018MNRAS.479.4786V}, where peaks at $\approx$1.5 and $\approx$2.0 Earth radii and a dip at $\approx$ 1.8 Earth radii were found. The interpretation of this radius valley is that it reflects a dichotomy in planetary properties: while the close-in planets with radii of less than 1.5 Earth radii are generally thought to be atmosphereless rocky planets, planets with radii larger than 2.0 Earth radii are thought to be mini-Neptunes with rocky cores and hydrogen--helium envelopes or ice-dominated planets (e.g. \citealt{2013ApJ...775..105O, 2014ApJ...792....1L, 2018ApJ...853..163J}). This indicates a compositional difference between these types of planets and may suggest a different origin. The radius valley in itself seems to be universal in the sense that it appears at all stellar masses \citep{2024MNRAS.531.3698H}.

The radius valley can be explained by photoevaporation of close-in planetary atmospheres (e.g. \citealt{2013ApJ...775..105O, 2014ApJ...792....1L}) or by core-powered mass loss (e.g. \citealt{2019MNRAS.487...24G}). Alternatively, atmospheres of planets could be lost during collisions with other planets (e.g. \citealt{2015ApJ...812..164L, 2016ApJ...817L..13I}). The current data support all of these scenarios and are in particular consistent with a combined scenario, where all these effects can act at the same time \citep{2022ApJ...939L..19I}.

As the planets migrate inwards, they will stop at the inner edge of the  disc \citep{2006ApJ...642..478M, 2019A&A...630A.147F}, where they can then build up resonance chains \citep{2017MNRAS.470.1750I, 2018A&A...615A..63O, 2019arXiv190208772I}. As the gas disc disappears, the damping forces acting on the planets diminish and the resonance chains can be broken by instabilities, which rearranges the planetary orbits and increases the eccentricities and inclinations of the close-in planets. These instabilities can either be self-triggered \citep{2017MNRAS.470.1750I, 2019arXiv190208772I} or can be triggered due to the presence of outer giant planets (e.g. \citealt{2023A&A...674A.178B}). In addition, the instabilities are affected by the proximity of the planets to each other (e.g. \citealt{1993Icar..106..247G}) as well as by their masses. The rearrangement of the planets within these systems can naturally explain the observed number of transiting planets, as well as the general period distribution of these planets \citep{2019arXiv190208772I}, even though the simulations have trouble explaining systems in the 2:1 resonance configuration.

The trapping of planets into MMRs depends on the relative migration speed of the planets (e.g. \citealt{2023MNRAS.522..828H}). Faster migrating planets can result in more compact systems than slower migrating planets. The migration velocity in itself, however, is set by the properties of the disc, such as its viscosity. At low viscosity, even mini-Neptunes can start to open partial gaps that can reduce their migration speed (e.g. \citealt{2011MNRAS.410..293P, 2012ARA&A..50..211K, 2013A&A...550A..52B}). Consequently, systems forming in low-viscosity environments could harbour wider resonances, with consequences for the long-term evolution of these planetary systems. On the other hand, a lower migration speed might imply that the planets are not trapped in resonance at all at the end of the gas-disc phase, because they migrate too slowly to reach the inner edge of the disc (e.g. \citealt{2017AJ....153..222B}).

Past simulations of the `breaking the chains' model focused on systems without giant planets in high-viscosity environments \citep{2017MNRAS.470.1750I, 2019arXiv190208772I, 2022ApJ...939L..19I, 2020MNRAS.497.2493E, 2022MNRAS.509.2856E, 2023MNRAS.521.5776E} or on systems with outer giants, but in low-viscosity environments (e.g. \citealt{2020A&A...643A..66B, 2023A&A...674A.178B}). The low-viscosity simulations with giant planets already seem to suggest that systems with a 2:1 resonance configuration could be more common in these environments. Here we want to bridge this gap by analysing simulations within the breaking the chains model at low disc viscosity without outer giant planets.

This paper is structured as follows. In Sect.~\ref{sec:methods} we give a short overview of our model. We show the results of our simulations in Sect.~\ref{sec:simulations} and compare these with the results of the high-viscosity simulations of \citet{2019arXiv190208772I}. We then discuss our results in the context of other works in Sect.~\ref{sec:discussion} and summarise our conclusions in Sect.~\ref{sec:summary}.

\section{Methods}
\label{sec:methods}

N-body simulations
are needed to study the dynamics and evolution of planetary systems. To this end, we employ FLINTSTONE, an N-body framework based on the MERCURY integrator \citep{1999MNRAS.304..793C} that includes planetary growth and migration \citep{2019A&A...623A..88B, 2019arXiv190208772I}. We use the same code here as that described in detail in \citet{2019arXiv190208772I} and \citet{2019A&A...623A..88B}, and only summarise the main ingredients of our simulations.

We initialise our simulations with 30 planetary embryos separated by 0.25 AU starting at 2.75 AU in a low-viscosity environment ($\alpha=10^{-4}$ for all our simulations). In contrast, the simulations by \citet{2019arXiv190208772I} used a viscosity of $\alpha=5.4 \times 10^{-3}$, which is about a factor of 50 more viscous. We note that the change of viscosity only influences the migration velocity of planets (see below and Appendix~\ref{app:migration}), while the other disc parameters (e.g. gas surface density and temperature) are kept the same for simplicity. The gas surface density and temperature thus evolve in time exactly in the same way for all our simulations. This approach, while artificial, allows a clean approach to understand how the viscosity influences the formation of planetary chains. This approach is the same as in our previous works \citep{2020A&A...643A..66B, 2023A&A...674A.178B}. The planetary embryos of around Moon mass start in an already evolved protoplanetary disc that is 2 Myr old in the framework of \citet{2015A&A...575A..28B}, as in our previous work \citep{2019A&A...623A..88B, 2020A&A...643A..66B, 2023A&A...674A.178B}. The embryos have an initial eccentricity and inclination that is randomly assigned ($e<0.01$ and $i<0.5$ degrees).

At the start of the simulations, the planetary embryos start to grow via pebble accretion (see \citealt{Johansen2017} for a review), where we follow the recipe of \citet{Johansen2015}. We use an exponentially decaying pebble flux that corresponds to a total of 350 Earth masses over the 3 Myr lifetime of the gas disc. The pebble sizes are limited by drift \citep{2014A&A...572A.107L} and we reduce the pebble flux by 50\% interior of the water ice line to account for water-ice evaporation. Furthermore, we then also reduce the pebbles sizes to millimetre(mm) size, in agreement with chondrules in the Solar System, as done in our previous work \citep{2019A&A...623A..88B, 2019arXiv190208772I}. Once the planets have reached the pebble isolation mass \citep{2014A&A...572A..35L, 2018arXiv180102341B, 2018A&A...615A.110A}, they stop growing by pebble accretion. 

Once planets have reached the pebble isolation mass, where we ignore the dependency on viscosity for
simplicity, they should start to accrete gas \citep{2014A&A...572A..35L, 2015A&A...582A.112B}. The exact efficiency of gas accretion is influenced by the planetary core mass as well as by the envelope opacity \citep{2000ApJ...537.1013I, 2019A&A...632A.118S, 2021A&A...647A..96B, 2021A&A...653A.103B, 2021A&A...646L..11M}. In addition, efficient atmospheric accretion can be hindered by recycling flows \citep{2017MNRAS.471.4662C, 2017A&A...606A.146L, 2021A&A...646L..11M, 2023MNRAS.523.6186W}, which transport material that entered the Hill sphere away from the planet before it can be accreted, which is especially efficient in the inner disc regions. As we are primarily interested in the formation of super-Earth/mini-Neptune systems, we chose to not model the gas accretion process onto the planets as in \citet{2019arXiv190208772I}. In principle, one can interpret this approach as a limit of very high envelope opacity that prevents efficient contraction of the planetary atmospheres, resulting in the formation of planetary cores with a small hydrogen/helium envelope.

As the planets start to grow, they eventually become big enough to start to migrate efficiently via type-I migration, where we follow the formulae from \citet{2011MNRAS.410..293P}. When the planets start to become even larger, they can open gaps in the protoplanetary disc, reducing their migration velocity, where we follow the approach of \citet{2018arXiv180511101K}, as discussed in Appendix~\ref{app:rates}. In addition, we apply forces to mimic the eccentricity and inclination damping onto the growing planets, following our previous approaches \citep{2019A&A...623A..88B, 2019arXiv190208772I}. In principle, the damping of eccentricity and inclination is affected by partial gap opening \citep{2023A&A...670A.148P, 2024ApJ...967..111P}, which results in a deviation from the damping formulae employed
in this work, which are from \citet{2008A&A...482..677C}. We nevertheless use the same damping formulae as used in previous works in the interest of consistency, and will investigate the influence of the damping formulae in future works. Planets in low-viscosity environments can open partial gaps, resulting in slower migration compared to planets in high-viscosity environments (see Appendix~\ref{app:migration}).

After 3 Myr, our protoplanetary disc dissipates on an exponential timescale, where we reduce the gas surface density uniformly in the disc with an exponential function on a timescale of 12 kyr within the last 100 kyr. Once the protoplanetary disc has dissipated, we integrate the simulations up to 100 Myr to allow sufficient time for instabilities to occur. During the instabilities, the planets can also collide. We treat collisions as perfect mergers between the planets, due to the fact that even more detailed collisional treatments that include fragmentation \citep{2012ApJ...745...79L} give the same results in our simulations \citep{2022MNRAS.509.2856E}.

At the end of the simulations, we investigate the structure of the planetary systems and also run synthetic observations of the formed systems. The exact method for this is described in \citet{2019arXiv190208772I}. These synthetic observations are necessary to effectively account for observational bias and compare our results with observations.

The initial conditions (eccentricity, inclination, and planetary mass) are all randomised. In the interest of statistical robustness, we ran 50 simulations with slightly varying initial conditions. We show the resulting planetary systems after 3 (end of the gas-disc phase) and 100 Myr of evolution in Appendix~\ref{app:systems}.

\section{Planetary systems from low-viscosity environments}
\label{sec:simulations}

In this section we discuss  the outcome of our results. We first investigate the general outcomes of our simulations and compare these to systems formed in the simulations by \citet{2019arXiv190208772I}. We then investigate how synthetic observations of these formed systems can explain the period distribution.

\subsection{Individual system evolution}

\begin{figure*}
 \centering
 \includegraphics[scale=0.8]{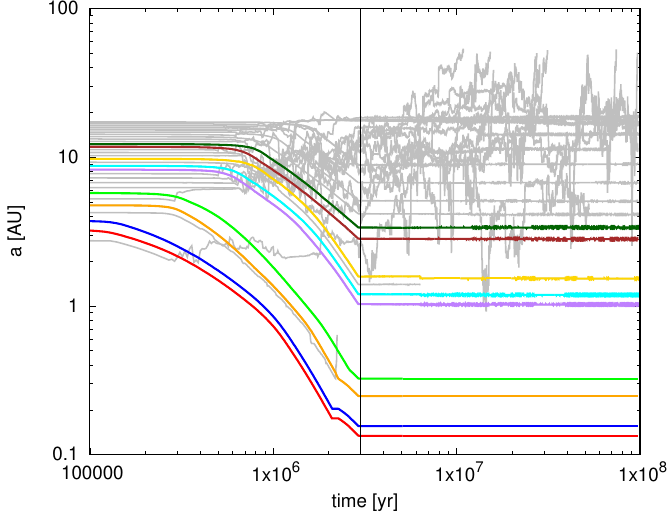}
 \includegraphics[scale=0.8]{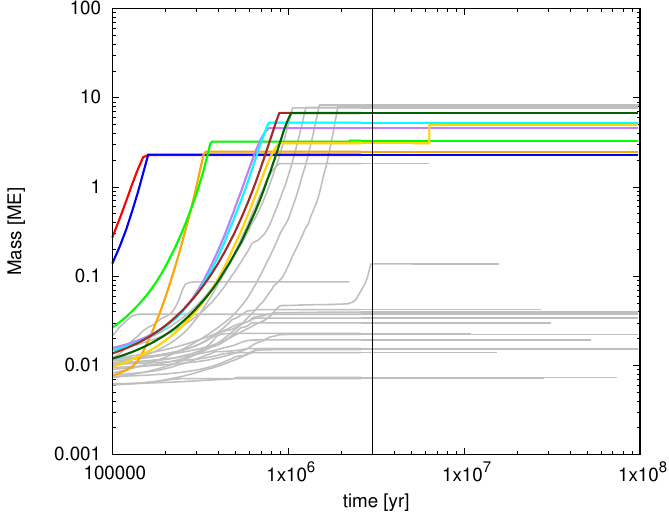} 
 \includegraphics[scale=0.8]{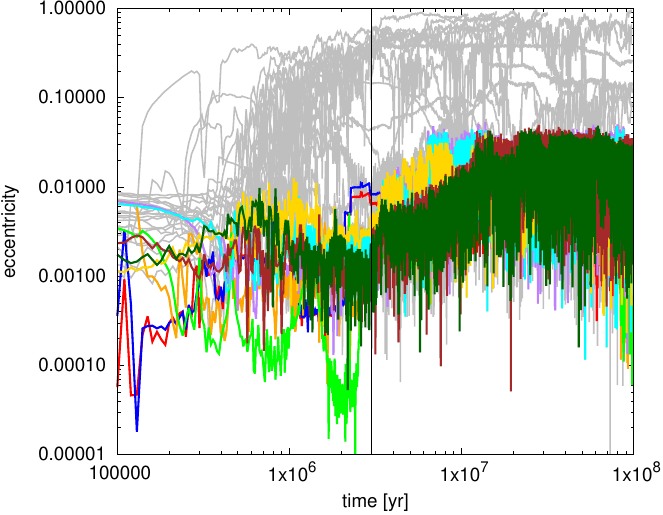}
 \includegraphics[scale=0.8]{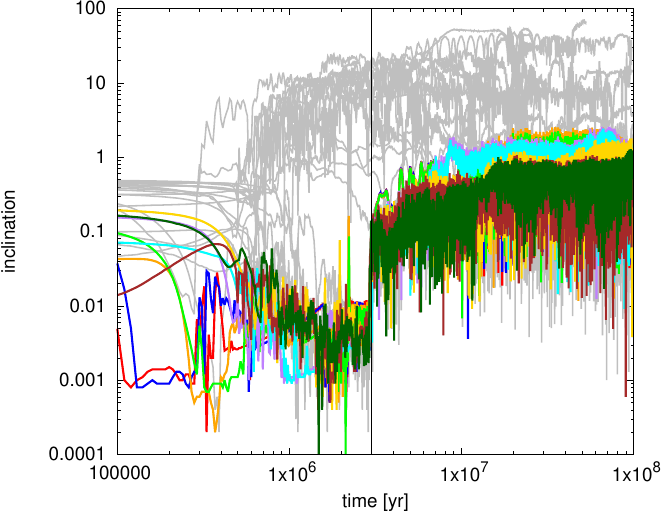}  
 \caption{Evolution of a single planetary system as a function of time regarding semi-major axis (top left), mass (top right), eccentricity (bottom left), and inclination (bottom right). The coloured lines correspond to the surviving inner (r<4 AU) planets, while the grey lines depict the evolution of the outer (also less massive) planets. The vertical line at 3 Myr marks the dissipation of the gas disc phase, where the planets lose the damping forces on eccentricity and inclination from the disc. The inner system remains stable until 100 Myr. The system corresponds to system number 3 in Fig.~\ref{fig:systems}.
   \label{fig:system3}
   }
\end{figure*}

 Figure~\ref{fig:system3} shows the time evolution of a system with respect to its evolution in semi-major axis, planetary masses, eccentricity, and inclination. The planets initially grow by accretion of pebbles, resulting in a smooth increase in the planetary masses. As the planets grow above 1 Earth mass, they start to migrate into the inner disc regions and get caught in resonance configurations. As the pebble-isolation mass increases with orbital distance (see Appendix~\ref{app:masses}), the planets forming further out in the disc become slightly more massive compared to the inner planets. During their migration, smaller planets can already be scattered.

Once the gas disc dissipates, the planetary eccentricities start to increase to a few percent, while the inclinations increase to around 1 degree for the massive inner planets. The eccentricities of the larger planets remain quite low, resulting in stable planetary systems for 100 Myr of evolution. Consequently a chain of resonant planets (5:4, 2:1 and 3:2 MMRs) survives in the inner disc.

\begin{figure*}
 \centering
 \includegraphics[scale=0.8]{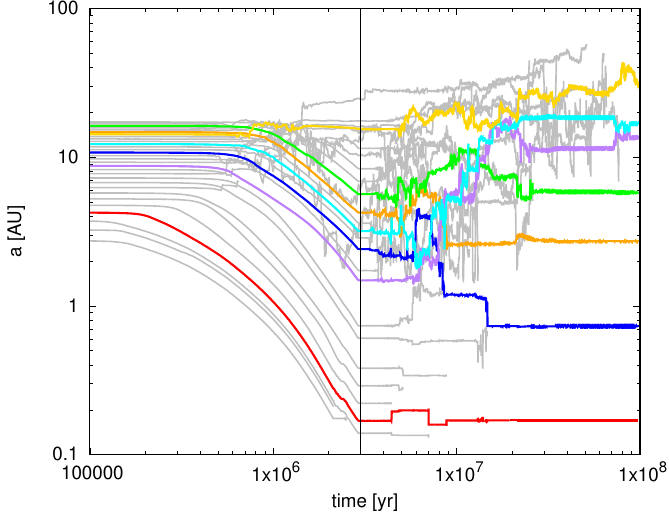}
 \includegraphics[scale=0.8]{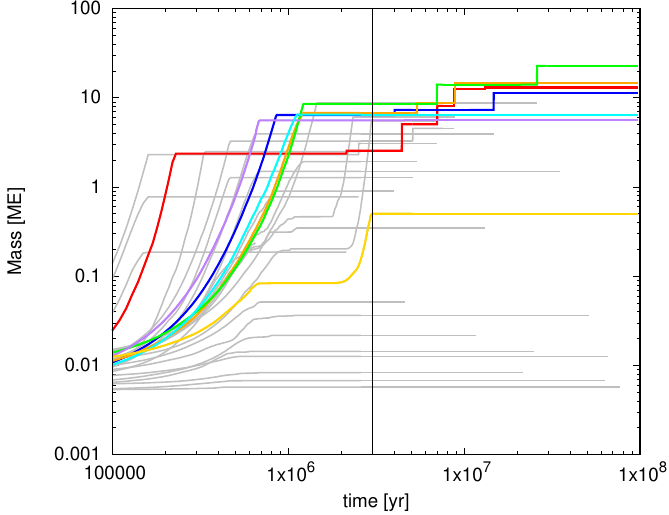} 
 \includegraphics[scale=0.8]{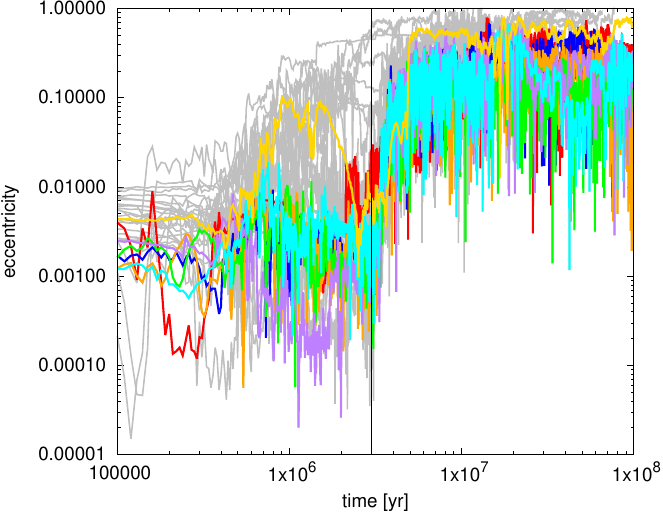}
 \includegraphics[scale=0.8]{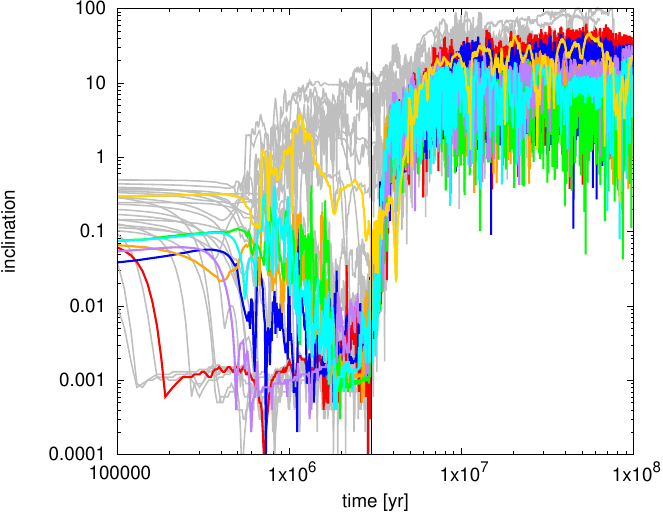}  
 \caption{Same as Fig.~\ref{fig:system3}, but for a system that became unstable after the gas-disc phase with two surviving inner planets. We mark all surviving planets in colour. The planet marked in yellow stops growing at around 700kyr, when it acquired an eccentricity too large to allow the accretion of pebbles. Once its eccentricity is damped below 1\%, it continues to grow via pebble accretion. The system corresponds to system number 21 in Fig.~\ref{fig:systems}.
   \label{fig:system21}
   }
\end{figure*}

\begin{figure*}
 \centering
 \includegraphics[scale=0.8]{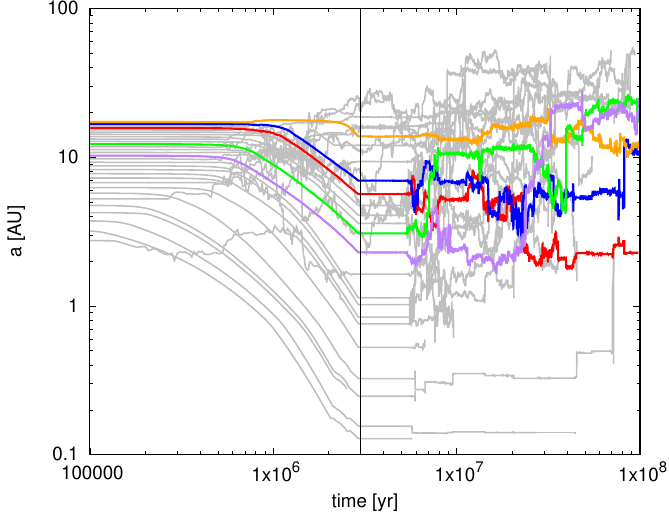}
 \includegraphics[scale=0.8]{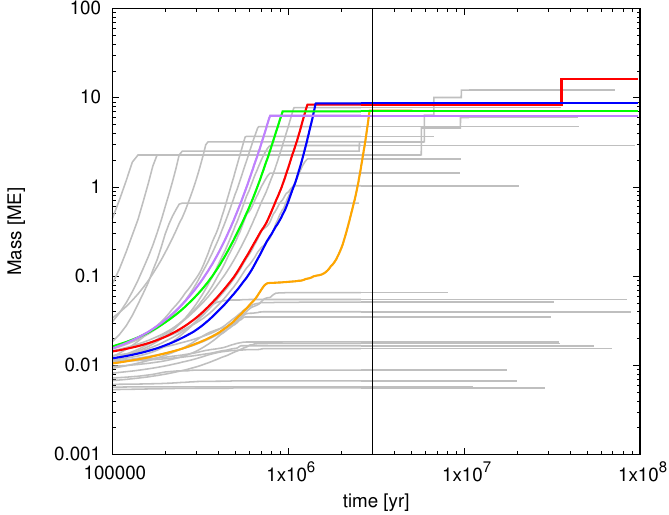} 
 \includegraphics[scale=0.8]{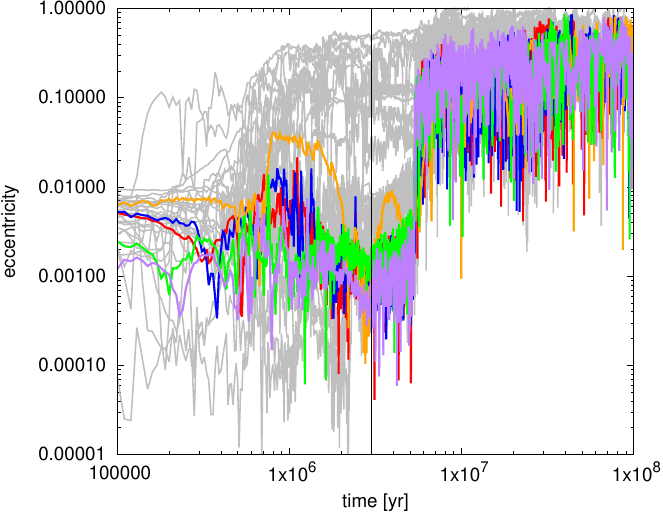}
 \includegraphics[scale=0.8]{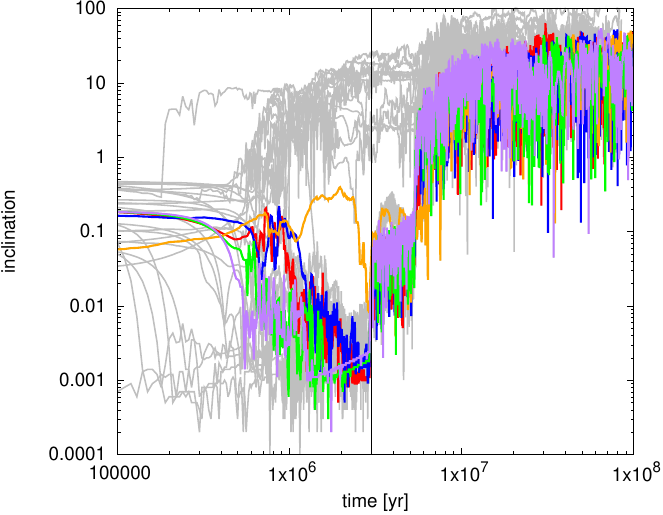}  
 \caption{Same as Fig.~\ref{fig:system3}, but for a system that became unstable after the gas-disc phase with no surviving inner planets. We mark all surviving planets in colour. The planet marked in orange acquires some eccentricity during the gas-disc phase, preventing pebble accretion, and it only continues to grow once its eccentricity is damped below 1\%. The system corresponds to system number 1 in Fig.~\ref{fig:systems}.
   \label{fig:system1}
   }
\end{figure*}

Figure~\ref{fig:system3} shows the evolution of a stable system, which survived until 100 Myr of evolution. Our simulations show two other typical outcomes: Fig.~\ref{fig:system21} shows a system that underwent instabilities after the gas-disc phase that sculpted the inner system, resulting in only two remaining inner planets, while Fig.~\ref{fig:system1} shows the evolution of a system where the inner system is completely removed by scattering caused by the outer planets.

The systems shown in Figs.~\ref{fig:system21} and \ref{fig:system1} experience strong scattering events after the removal of the gas discs, which leads to large increases in the planetary eccentricities to values of around 10\% (i.e. around ten times larger than the stable system in Fig.~\ref{fig:system3}).
This is a consequence of the dynamical instabilities in the outer parts of the system ($r>1$ AU), which eventually affect the inner system ($<0.5$ AU). Consequently, only two planets survive outside of any resonance configuration interior to 1 AU  (Fig.~\ref{fig:system21}). On the other hand, the instabilities for the system shown in Fig.~\ref{fig:system1} are so strong that the whole inner system is removed. However, the cause of the instabilities is the same: an increase in the orbital eccentricities in the outer system after the gas disc dispersal, leading to an instability that affects and removes the inner system.

\subsection{Planetary systems directly after the gas disc phase}

\begin{figure*}
 \centering
 \includegraphics[scale=0.53]{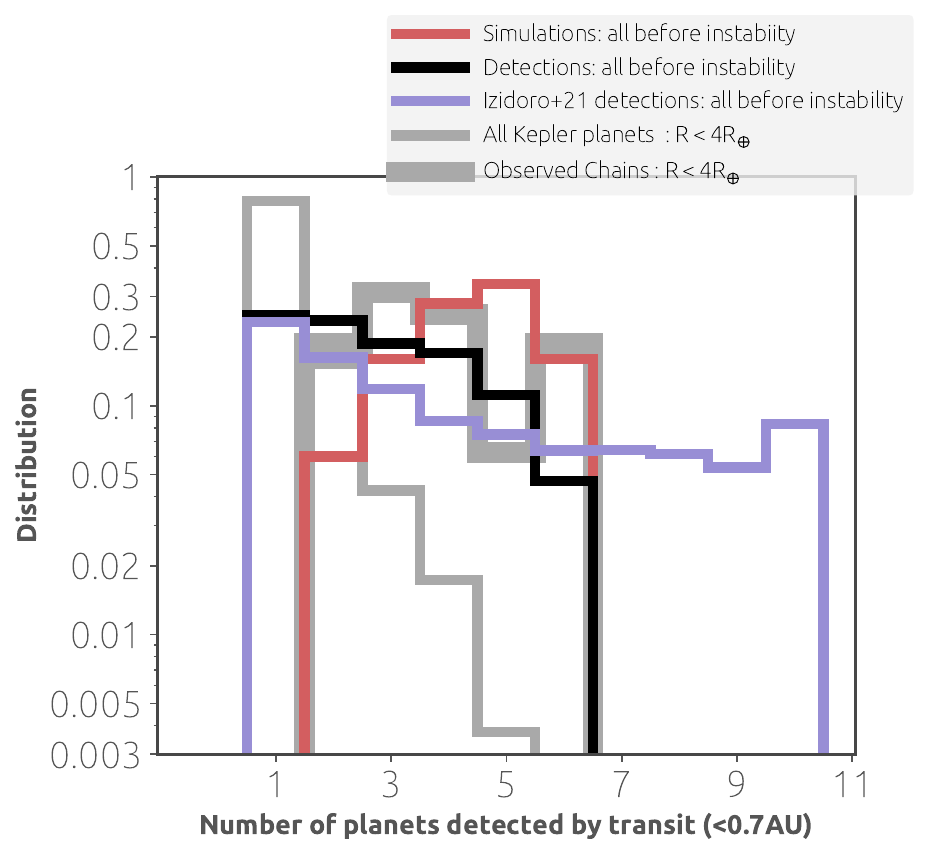}
 \includegraphics[scale=0.53]{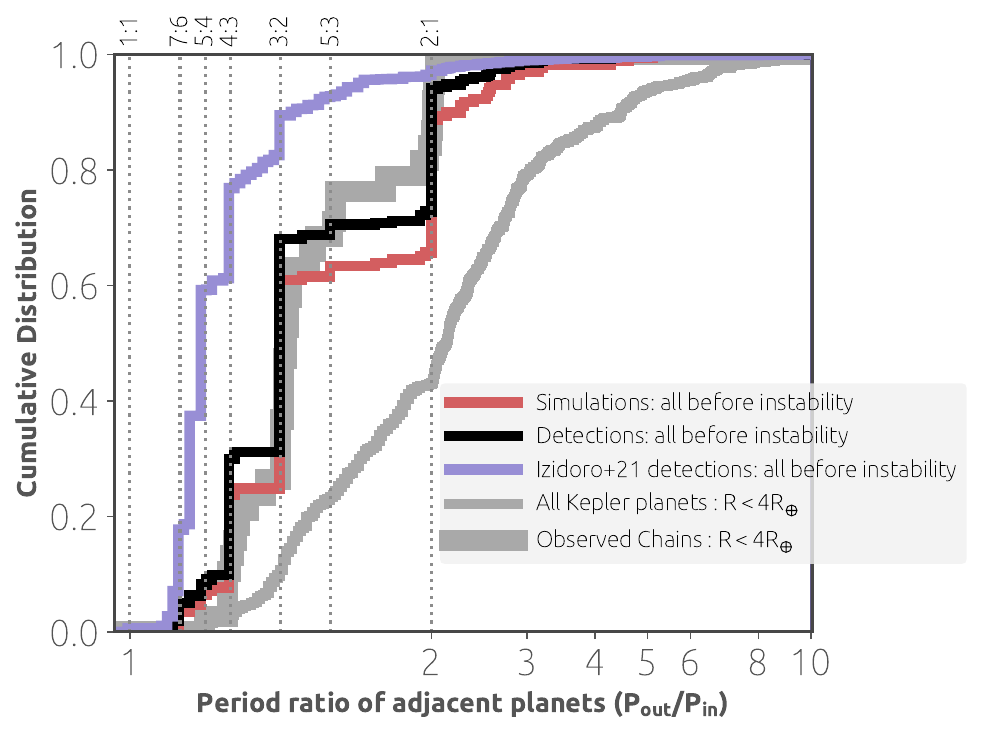} 
 \caption{Number of synthetically observed planets (left) and period ratios of adjacent planets (right) for the planetary systems directly at the very end of the gas-disc phase, before the systems undergo instabilities. The thin grey line marks all the observations, while the thick grey line marks the observed chains of planets. The red colour corresponds to the direct results of our simulations, while the black line corresponds to the synthetical observations of our simulations. The purple line represents the synthetic observations from \citet{2019arXiv190208772I}. A clear difference is noted in the compactness of the chains between the high-viscosity simulations of \citet{2019arXiv190208772I} and the low-viscosity simulations used  here.
   \label{fig:beforeInstability}
   }
\end{figure*}

During the first 3 Myr of our simulations, the planets can grow via pebble accretion (and collisions) and migrate, while simultaneously their eccentricity and inclination are damped by gas--disc interactions. Once the gas disc dissipates, only gravitational interactions remain, leading to an increase in eccentricity and inclination with the consequence that instabilities will occur. However, the structure of the systems after the gas-disc phase can already provide some initial hints as to the differences between low- and high-viscosity planet formation environments.

Fig.~\ref{fig:beforeInstability} shows the number of planets that have formed at the end of the gas-disc phase (left), as well as their period ratios (right). We also show the systems that could be identified by observations as well as the synthetic observations of the simulations of \citet{2019arXiv190208772I}. In addition, we show the observed \textit{Kepler} sample (thin grey lines) as well as the planetary chains observed by \textit{Kepler} (thick grey lines). We note that we only consider planets interior to 0.7 AU as observable, even though more planets can be hidden exterior to that (see Appendix~\ref{app:systems}). In addition, we only include planets above 1 Earth mass in our statistical analysis, because smaller planets are very likely to be missed in transit observations (e.g. \citealt{2018AJ....156...24M})

Within 0.7 AU, our low-viscosity simulations show a maximum of six planets, while the high-viscosity simulations of \citet{2019arXiv190208772I} show systems with up to ten planets. This difference can be explained by the slightly different setups used here. In particular, \citet{2019arXiv190208772I} used 60 planetary embryos per simulation, allowing more planets to be present in the first place. Another difference is the migration velocity of the growing planets, which is determined by the disc viscosity. Discs with lower viscosity allow early (partial) gap opening, which results in a transition of the planets into a slower migration regime compared to their high-viscosity counterparts (see Appendix~\ref{app:migration}). Consequently, not as many planets from the outer disc migrate interior to 0.7 AU, resulting in different systems.

While the systems in our simulations always have at least two planets with 1 Earth mass or more, observationally, the picture is different. Due to the mutual inclinations between the planets, nearly 30\% of all systems would only show one transiting planet, while the number of systems with five transiting planets is significantly reduced compared to the pure simulation data. This is in line with the results of \citet{2019arXiv190208772I}, who observed the same behaviour in their simulations. Comparing to the observational constraints from \textit{Kepler} clearly shows that the number of single transiting planets is not reproduced at the stage directly after dissipation of the  gas disc; this requires some instability to increase the level of mutual inclination and to reduce the number of planets (see below).

The period ratios on the other hand show that most of the formed planets are in resonant configurations at the end of the gas-disc lifetime, which is caused by inward migration and trapping at the  inner edge of the disc. This is, of course, in contrast to the observed period ratios by \textit{Kepler}, where only a small fraction of systems are expected to be in a resonant configuration. This result is in line with the findings of \citet{2019arXiv190208772I}, where instabilities are necessary to break the resonant configuration; we study this aspect further  below. However, there is a significant difference in the observed distribution of the period ratios between the simulations presented here and those of \citet{2019arXiv190208772I}: the resonant chains in \citet{2019arXiv190208772I} are much more compact than those in the simulations presented here.

The difference in the period ratios of the systems can again be explained by planetary migration. As the planets migrate faster in the \citet{2019arXiv190208772I} simulations, they can naturally come closer to each other than they are able to in simulations in a low-viscosity environment. In particular, the simulations by \citet{2019arXiv190208772I} seem to overestimate planetary systems in the 7:6 and 5:4 MMRs, while simultaneously underestimating the planets in the 3:2 and 2:1 MMRs. In contrast, the low-viscosity simulations show a much better match at these MMRs. We show in Appendix~\ref{app:number} that this difference is truly related to disc viscosity rather than the number of embryos. On the other hand, here we only discuss the simulations at the end of the gas-disc lifetime, and instabilities are expected to change these configurations. We discuss this aspect further in the following section.

\subsection{Planetary systems after the instability phase}

After the gas phase, the planets can only interact gravitationally and are no longer influenced by the damping effects of the disc that previously acted on their eccentricity and inclination. Consequently, the planetary systems become unstable and their orbital configurations  change. In the following, we classify systems that have not undergone a major rearrangement after the gas-disc phase as `{chains}'.

\begin{figure*}
 \centering
 \includegraphics[scale=0.55]{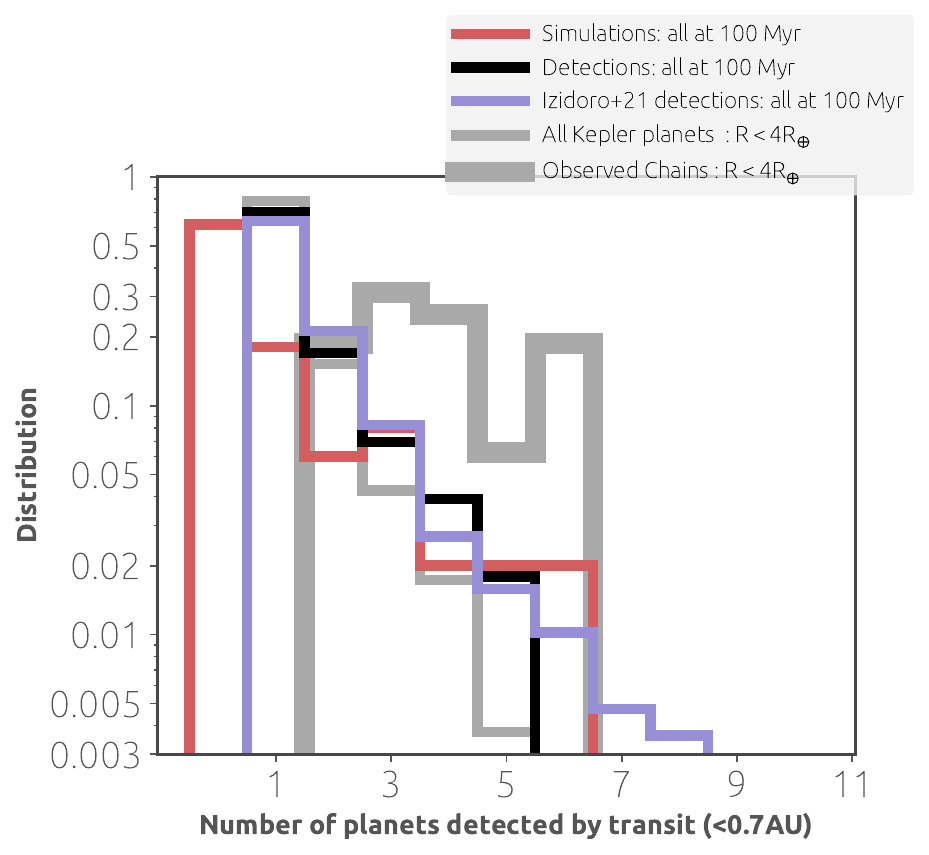}
 \includegraphics[scale=0.55]{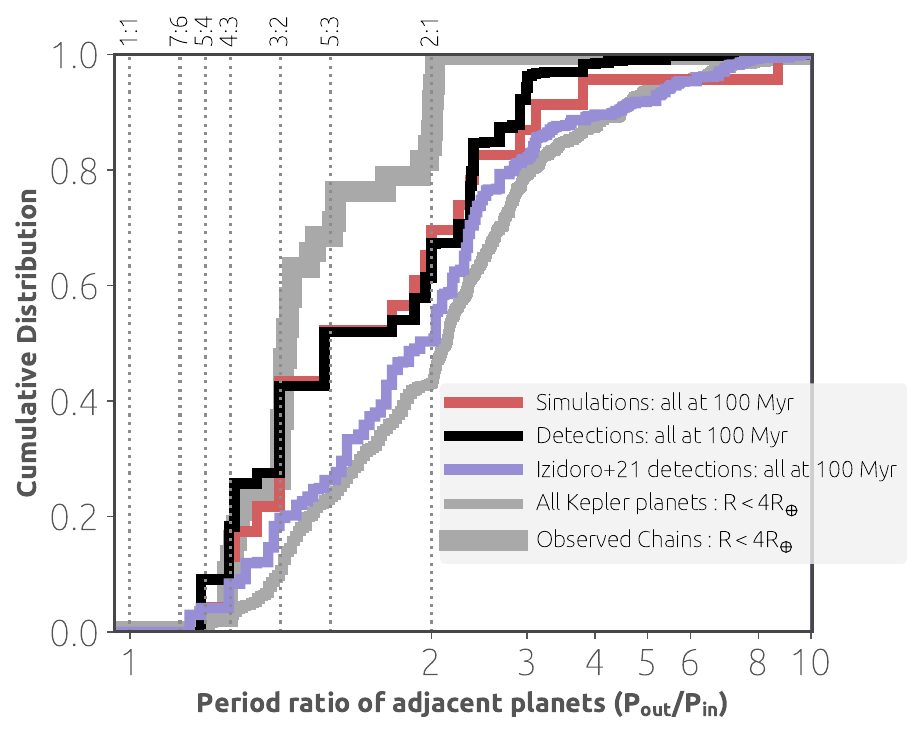} 
 \caption{Number of synthetically observed planets (left) and period ratios of adjacent planets (right) for the planetary systems after 100 Myr of integration; i.e. after the instability phase. The colours are the same as in Fig.~\ref{fig:beforeInstability}. The high-viscosity simulations from \citet{2019arXiv190208772I} produce wider systems after the instabilities compared to the low-viscosity simulations, which is also a consequence of the real number of planets in the systems before instabilities (left in Fig.~\ref{fig:beforeInstability}), where more planets are initially present at high viscosities.
   \label{fig:allafter}
   }
\end{figure*}

Fig.~\ref{fig:allafter} shows the number of detected planets (left) and their period ratios (right) in our simulations after 100 Myr of system evolution. As before, we also show the distribution of the detected systems as well as the results from \citet{2019arXiv190208772I} in comparison to the \textit{Kepler} data. 

As a consequence of the instability, many planetary systems have no inner system left and therefore show no inner planets (see Appendix~\ref{app:systems}). Consequently, the number of systems with three or more inner planets is also greatly reduced. However, the reduced number and increased mutual inclination of the systems allow a much better match to the \textit{Kepler} observations regarding the systems with only one transiting planet, in line with the results of \citet{2019arXiv190208772I}. The systems with multiple transiting planets also match much more closely after the instability phase.

The resulting period ratios are determined by the systems with at least two surviving planets (limiting our statistics to some level; see Appendix~\ref{app:systems}). The general results of our low-viscosity simulations do not match the period distribution from the \textit{Kepler} systems, as the systems in our simulations remain too tightly packed compared to the \textit{Kepler} systems, even when applying synthetic observations. In contrast, the simulations of \citet{2019arXiv190208772I} are a much better match to the \textit{Kepler} systems in terms of their period ratio distribution.

The differences in these results could be explained by the different strengths of the instabilities between the different system configurations. While the instabilities in our simulations mostly result in a removal of planets from the system, the instabilities in the simulations of \citet{2019arXiv190208772I} mostly show a rearrangement of the system, with more survivors, which leads to period ratio distributions that are in better agreement with observations. This is caused by the difference in the outer system architecture. While the systems formed in the high-viscosity environments show almost no massive planets exterior to $\sim$200 days due to the fast inward migration of planets of a few Earth masses, the opposite is true at low viscosity, where even the more massive planets cannot migrate to the inner disc (see Appendix~\ref{app:migration}), resulting in outer systems with several planets above a few Earth masses. In combination with the lower planetary masses of the inner systems in the low-viscosity case (see Appendix~\ref{app:masses}), the instabilities triggered by the outer systems are more likely to result in the removal of the inner planets rather than simply a dynamical rearrangement. Using a mixture of 2\% stable and 98\% unstable systems (as done in \citet{2019arXiv190208772I} to achieve the best match to observations) does not allow a good match to the overall period ratios (see Appendix~\ref{app:periods}) of the low-viscosity simulations as well.

\begin{figure*}
 \centering
 \includegraphics[scale=0.55]{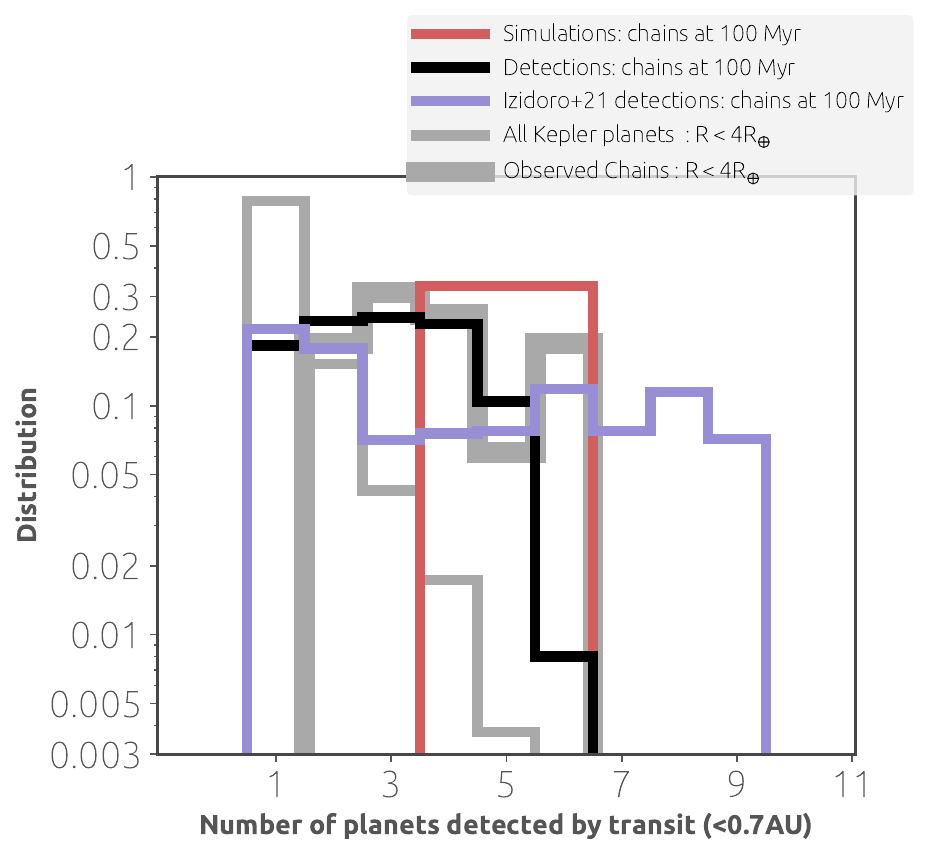}
 \includegraphics[scale=0.55]{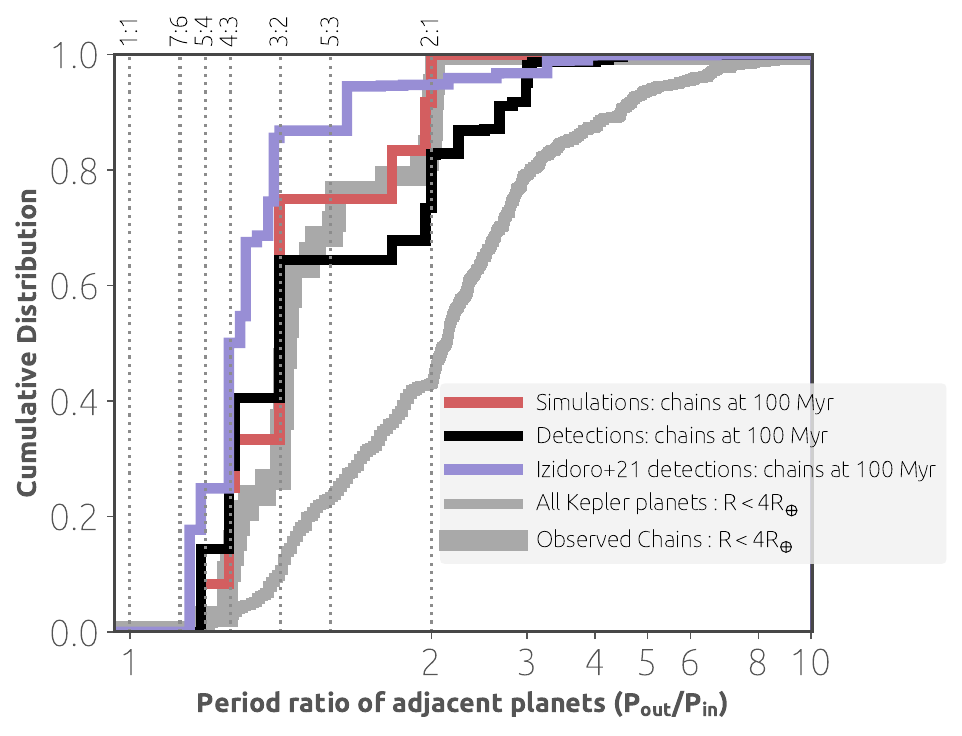} 
 \caption{Number of synthetically observed planets (left) and period ratios of adjacent planets (right) for chains of planets that are stable after 100 Myr. The colours are the same as in Fig.~\ref{fig:beforeInstability}.
   \label{fig:chainsafter}
   }
\end{figure*}

In Fig.~\ref{fig:chainsafter} we plot the number of observed planets and their period distributions {only} for the chains formed in our simulations, as well as the chains from \citet{2019arXiv190208772I} after 100 Myr of evolution. The chains from our simulations harbour only four or more planets, in contrast to the \textit{Kepler} observations; this is also different from the simulations of \citet{2019arXiv190208772I}, who also show chains with a smaller number of planets. In addition, the high-viscosity simulations predict chains with seven or more planets, which have not yet been detected in such abundance, but should exist in theory (e.g. Trappist-1).

On the other hand, the surviving chains show a wider period ratio distribution in the low-viscosity simulations than in the high-viscosity simulations. In particular, the low-viscosity simulations show more planets in the 3:2 and 2:1 configurations compared to the high-viscosity simulations, which leads to better agreement with the observational constraints from the chains alone.

\begin{figure*}
 \centering
 \includegraphics[scale=0.6]{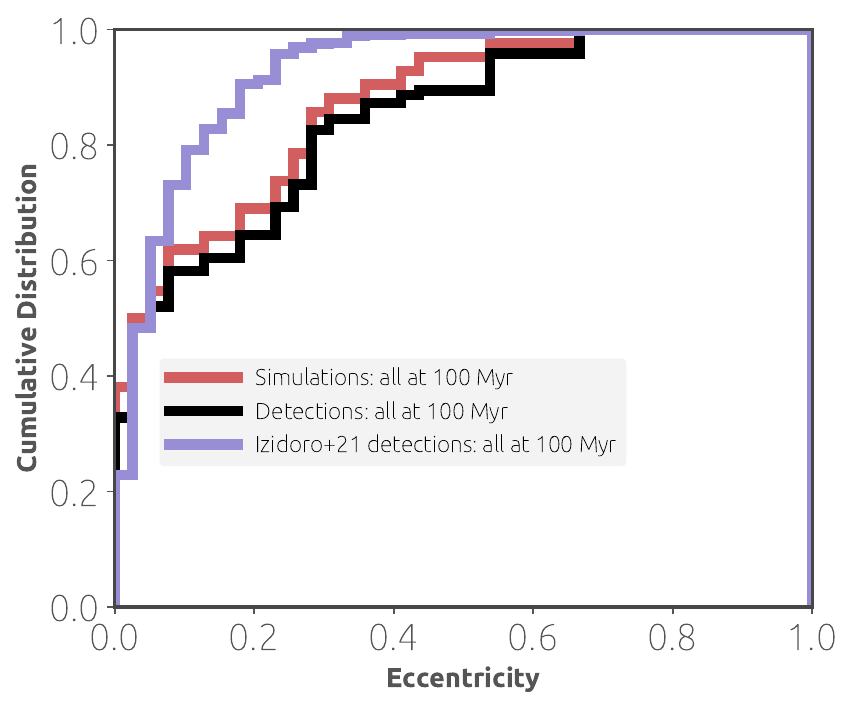}
 \includegraphics[scale=0.6]{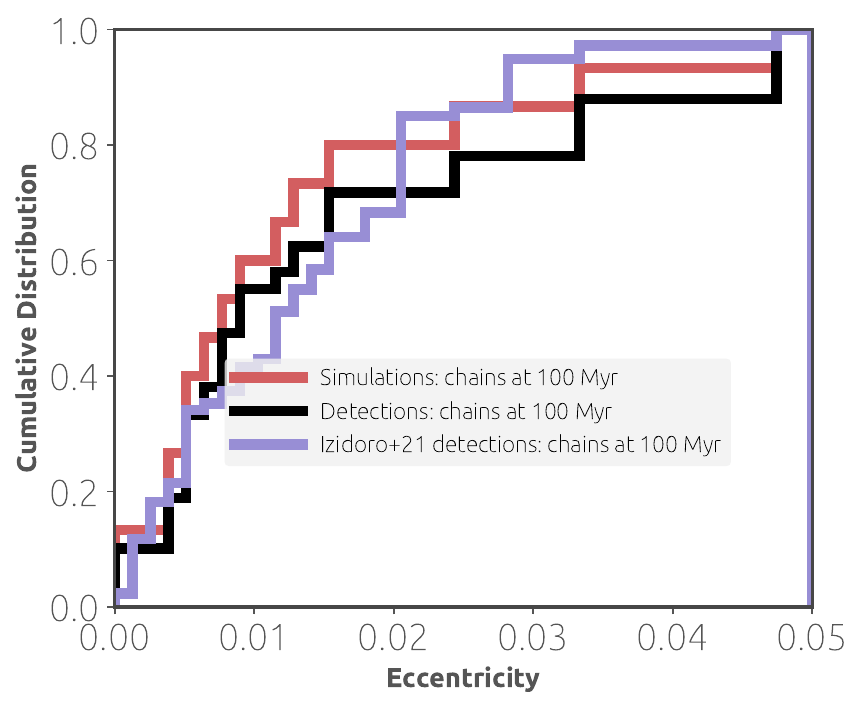} 
 \caption{Eccentricity distribution of our low-viscosity simulations and of the high-viscosity simulations of \citet{2019arXiv190208772I} for all planets (left) and only for the surviving chains (right). 
   \label{fig:eccentricity}
   }
\end{figure*}

The eccentricities of the planets are shaped by the gravitational interactions between the bodies, where more violent instabilities can cause larger eccentricities. Consequently, the eccentricity distributions of the two sets of simulations (Fig.~\ref{fig:eccentricity}) are different. The low-viscosity simulations experience more violent instabilities, which are caused by the outer planets, while the higher viscosity simulations do not feature outer planets that are
both sufficiently massive and sufficiently close to the inner system to affect it. Consequently, they experience only `mild' instabilities, resulting in planets with  larger eccentricities in the low-viscosity simulations.

In contrast, the eccentricity distribution of the surviving planets in chains (right in Fig.~\ref{fig:eccentricity}) is very similar. This is caused by the fact that chains, if they are to survive, must also somehow avoid being strongly affected by external planets and must therefore keep their eccentricities low. In the planetary systems shown in Fig.~\ref{fig:system1} and Fig.~\ref{fig:system21}, the eccentricities increase above 10\%, resulting in violent instabilities that break the chains. In contrast, systems that remain in a resonant configuration have low eccentricities (Fig.~\ref{fig:system3}), as the low eccentricities prevent orbit crossing and instabilities in general.

We note that, in principle, more compact resonance configurations (e.g. 7:6, 6:5, 5:4) should result in more violent instabilities compared to wider resonance configurations (e.g. 3:2, 2:1). An example of a well-described less violent instability would be the dynamical instability of the Solar System in the Nice model, although it features more massive planets on much larger orbits (e.g. \citealt{2005Natur.435..459T, 2018ARA&A..56..137N}). In the cases analysed here, instabilities tend to be more violent, not necessarily due to the compactness of the resonant configurations themselves, but instead due to the effect of external perturbers (see Fig.~\ref{fig:system1} and Fig.~\ref{fig:system21}). We will explore this mechanism in more detail in future work.

\section{Discussion}
\label{sec:discussion}

We show that the outcome of our simulations depends crucially on the viscosity of the disc (see Appendix~\ref{app:number}). We now discuss the influence of the different (changed) parameters on the outcomes of our simulations.

\subsection{Planetary migration}

The planetary migration rates are not only set by the viscosity of the disc (e.g. \citealt{2011MNRAS.410..293P}), but also by its other properties, and in particular its surface density and temperature (which in turn set the entropy). Here we use the migration formulae by \citet{2011MNRAS.410..293P}, which allow reduced (or even outward) migration rates at higher viscosities, depending on the disc profile. We also use the disc profile of \citet{2015A&A...575A..28B}, where the disc has already evolved for 2 Myr. At this stage, the aspect ratio of the  inner disc is relatively small (H/r $\approx$0.025-0.03), resulting in an easier partial gap opening by small planets \citep{2007MNRAS.377.1324C, 2013A&A...550A..52B, 2018arXiv180511101K, 2023A&A...670A.148P}, which reduces their inward migration speed compared to simulations at higher viscosity (see Appendix~\ref{app:migration}), where gap opening is hindered. Consequently, the exact outcomes of our simulations are influenced by our choice of  disc profile. 

We also note that other migration formulae (e.g. \citealt{2017MNRAS.471.4917J}) give different migration rates, where the heating torque in particular can change the migration direction of fast-growing small-mass planets (e.g. \citealt{2015Natur.520...63B, 2017MNRAS.472.4204M}). However, the heating torque diminishes once the planets reach the pebble-isolation mass, resulting in nominal inward migration at low viscosity. The overall outcome of planet-formation simulations is not greatly influenced by the heating torque or different migration prescriptions \citep{2020arXiv200400874B}; however, the detailed migration rates and disc parameters are important in the formation of specific systems (e.g. \citealt{2021A&A...656A.115H}).

The relative migration velocities of the planets set the resonances in which planets can be trapped (e.g. \citealt{2023MNRAS.522..828H}). This is independent of the {factor that sets}  the migration rates, be it disc profile or viscosity. Here we explore how low viscosities influence the migration rates. A similar effect could also be achieved via different disc parameters, such as a lower overall disc surface density. However, this would also influence the  rate at which planets grow via pebble accretion, as well as the damping rates of eccentricity and inclination, making it more difficult to make direct comparisons and draw robust conclusions.

In order to test the influence of disc viscosity on resonance trapping, we show in Appendix~\ref{app:2planets} the outcomes of simulations of two migrating equal-mass planets in different disc-viscosity environments. While the exact outcomes of the simulations depend on the disc parameters, it is clear that the viscosity ---which sets the migration rate--- determines the resonance that the planets are trapped in. Planets migrating in high-viscosity environments are trapped in closer resonances compared to planets migrating in low-viscosity environments.

\subsection{Planetary growth}

The growth of the planets is initially driven by pebble accretion (e.g. \citealt{2010MNRAS.404..475J, 2010A&A...520A..43O, 2012A&A...544A..32L}). While pebble accretion is an efficient mechanism of planet growth, individual pebble-accreting planets are actually very inefficient in reducing the overall inward flux of pebbles \citep{2014A&A...572A..35L, 2019A&A...623A..88B}, as long as they do not reach the pebble isolation mass. As the pebble flux is not reduced significantly by one growing planet, systems of multiple planets can efficiently form \citep{2019A&A...627A..83L, 2019A&A...623A..88B, 2019arXiv190208772I, 2021A&A...650A.116M}. In addition, the growth in the outer regions of the disc happens on longer timescales, meaning that the planets forming in the inner system reach the pebble isolation mass more quickly than those forming in the outer regions, diminishing the influence of more outer embryos on the growth of inner embryos \citep{2019A&A...623A..88B, 2020A&A...643A..66B, 2023A&A...674A.178B}.

In our simulations, we used 30 planetary embryos, while the simulations of \citet{2019arXiv190208772I} used initially 60 planetary embryos. Consequently, more planets can initially grow in the simulations of \citet{2019arXiv190208772I}, but they are also distributed in a wider radial space compared to our simulations. Furthermore, planets forming in the outer regions (e.g. exterior to 10 AU) will not migrate interior to 0.7 AU in low-viscosity environments (see Appendix~\ref{app:migration}), and remain in the outer disc, causing late instabilities of the inner systems. Consequently, we think that a greater number of embryos in the low-viscosity simulations would not significantly change the outcome of our simulations with respect to the inner system structure (see also Appendix~\ref{app:number}).

On the other hand, in the high-viscosity simulations of \citet{2019arXiv190208772I}, planets migrate inwards faster and can thus reach the inner edge of the disc even if they originate from beyond 10 AU. Consequently, more planetary embryos can migrate into the inner region of the disc. This can result in a squeeze of the formed planetary chains, where incoming planets can push already trapped planets into tighter resonances. In combination with the faster migration speed, which can allow tighter resonances as well, the formed systems are more tightly packed, with more planets in the high-viscosity simulations. 

In Appendix~\ref{app:number}, we show a comparison to the simulations of \citet{2023A&A...674A.178B}, who also performed simulations with 15 and 60 initial embryos. These simulations support the idea that the migration velocity ---set by disc viscosity--- is responsible for the different chain configurations rather than the number of planetary embryos.  In order to verify this result, we conducted simulations with only two planets in discs with different viscosities in Appendix~\ref{app:2planets}, confirming this phenomenon in
simulations very clearly.

\subsection{Formation of Kepler-223 and TOI-178}

The Kepler-223 system consists of four super-Earths locked in a resonance chain with 4:3, 3:2, and 4:3 resonances \citep{2016Natur.533..509M}. Understanding the formation of a particular resonance chain requires detailed studies of the migration history of the planets. \citet{2021A&A...656A.115H} carried out an extensive parameter study using N-body simulations with prescribed migration rates to explain the migration history of Kepler-223. In particular, these authors varied the gas surface density value and the disc viscosity in order to find a parameter combination that allows the formation of that system.

\citet{2021A&A...656A.115H} find that surface densities higher than the Minimum Mass Solar Nebular, or MMSN \citep{1977Ap&SS..51..153W, 1981PThPS..70...35H} (which are also higher than the gas surface densities used here) and $\alpha$-viscosities with $\alpha \le 10^{-3}$ favour the formation of the orbital configuration of Kepler-223. This result is in line with our low-viscosity simulations, which can reproduce a significant fraction of systems with 4:3 and 3:2 resonances. This result is encouraging considering the completely different setup of the simulations of \citet{2021A&A...656A.115H} and ours, indicating that lower viscosities could favour the production of systems in the 4:3 and 3:2 resonances. 

The TOI-178 system consists of six transiting planets following a 2:4:6:9:12 chain of Laplace resonances \citep{2021A&A...649A..26L}. Our simulations indicate that such wide chains could be better explained by low-viscosity environments, as the chains built in high-viscosity environments do not produce planets in the 2:1 resonance configuration.

\subsection{Consequences for inner and outer system structure}

Our simulations show that the structure of the inner system is influenced by the outer regions (P$>$200 days). While the systems that form in the low-viscosity environments experience dramatic instabilities that can result in a large fraction of completely destroyed systems, the instabilities in the high-viscosity systems mostly simply break the chains but leave the overall systems intact. We note that these instabilities do not necessarily eject the planets from the system; indeed, a large fraction of the planets are thrown into the host star, where they can influence the stellar atmospheric abundances (e.g. \citealt{2024Natur.627..501L}). This difference is caused by the greater number and larger masses of planets in the outer disc regions (exterior to 1 AU; see Fig.~\ref{fig:systems}), which trigger the violent instabilities in the systems formed in the low-viscosity environments. This is in line with the simulations of \citet{2023A&A...674A.178B}, where outer giant planets destroy the inner systems, with the difference that the systems in \citet{2023A&A...674A.178B} already become unstable during the gas-disc phase due to the greater impact of the giants compared to that of the outer sub-Neptunes in the simulations presented  here.

This difference has two important implications for the occurrence rates of planets in the inner systems as well as in the outer systems. Even though all our simulations allow the formation of inner super-Earths during the gas-disc stage, only around 50\% of our final systems harbour inner planets after 100 Myr of integration. On the other hand, all our systems still harbour planets exterior to 1 AU after 100 Myr of integration. This result is broadly in line with the occurrence rate of sub-Neptunes exterior to the water-ice line from microlensing surveys (e.g. \citealt{2016ApJ...833..145S}), which predict that these planets should be more common than hot inner super-Earths. 

We note that the presence of the outer systems (r>1 AU) is influenced by our choice of the initial embryo distribution. If we were to include only a small number of embryos in our simulations (e.g. less than 8) within a few astronomical units (AU), all embryos would migrate into the inner disc regions and no outer system would exist. In this case, the instabilities would be less pronounced and the inner systems would survive, as they do in the high-viscosity simulations. However, the lower viscosity would still determine the formed resonance chains, because the lower viscosity allows slower migration speeds, trapping the planets in wider configurations. This can be seen in Fig.~\ref{fig:beforeInstability}, where the period ratios of planets clearly show wider systems for the low-viscosity simulations compared to the high-viscosity simulations. On the other hand, if no outer embryos were to exist initially, it is unclear how the sub-Neptunes exterior to the water-ice line, as found in microlensing surveys (e.g. \citealt{2016ApJ...833..145S}), would come to exist.

\subsection{Comparison to other planet-formation studies}

The formation of super-Earth and mini-Neptune systems has been studied by various groups in general (e.g. \citealt{2018A&A...615A..63O, 2021A&A...656A..69E, 2024ApJ...972..181O, 2023NatAs.tmp...10B}) and specific systems have also been examined (e.g. \citealt{2019A&A...631A...7C}). These studies differ in their methods regarding accretion (pebbles or planetesimals), disc evolution (viscous or disc winds), and initial embryo formation (in ring-like structures or distributed all over the disc), but they all include planetary migration, which is necessary for the build up of resonance chains. More notably, most studies use the migration prescription by \citet{2011MNRAS.410..293P}, as we do in the work presented  here. Furthermore, all simulations include an inner edge, where inward migration can stop and planets can pile up (e.g. \citealt{2006ApJ...642..478M, 2019A&A...630A.147F}).

However, independently of how the planets form (pebbles or planetesimals) or  the configuration in which they form (in rings or widely spread), they have to migrate to the inner edge of the protoplanetary disc. We believe migration speed, which is determined by the viscosity of the gas disc (as well as by the gas surface density and temperature; see Appendix~\ref{app:rates}), to be the crucial component that determines the eventual resonance configuration of the planets. This belief is motivated by the fact that capture into resonances is determined by the relative migration velocity and not by the initial distance between the embryos.

On the other hand, as larger viscosities allow the inward migration of planets from beyond the water-ice line, a mixture in the composition of the systems is expected (e.g. \citealt{2022ApJ...939L..19I}). At low viscosity, this may not be the case, unless the discs are very cold and the water-ice line is initially close to 1 AU (e.g. \citealt{2019A&A...624A.109B}), allowing planets to reach the inner disc even in low-viscosity environments (see also Appendix~\ref{app:migration}). However, how much of the water ice in pebbles can be retained by growing planets is still debated, as the pebbles might evaporate in the hot atmosphere of the planets, preventing the efficient accretion of water ice \citep{2021SciA....7..444J, 2024A&A...688A.139M}.

\section{Summary}
\label{sec:summary}

We conducted simulations similar to those of \citet{2019arXiv190208772I}, but using a low-viscosity environment. This low-viscosity environment allows planets of a few Earth masses to already open partial gaps, reducing their  rate of inward migration. Consequently, the planetary chains that form at the inner edge of the disc are in wider configurations compared to planets that form in a high-viscosity environment. This has consequences for the stability of the systems and for their observability. Our findings are as follows:

\begin{itemize}
 \item The lower viscosity causes slower migration speeds and consequently planets remain in wider configurations after the gas-disc phase, where planets are mostly trapped in the 4:3, 3:2, and 2:1 resonances. The higher-viscosity simulations, on the other hand, do not populate the 2:1 resonance directly after the gas-disc phase and only marginally populate the 3:2 resonance. Most planets are in the 4:3 resonance or exhibit an even tighter configuration.
 
 \item Following  dissipation of the gas disc, the planetary systems experience instabilities, which break the resonance chains. Consequently, the number of observed planets can be matched, as  also shown in our previous works \citep{2019arXiv190208772I, 2023A&A...674A.178B}. 
 
 \item The overall period ratios of the low-viscosity simulations, after instabilities, remain too narrow to explain the observed period ratios. In contrast, the high-viscosity simulations, in general, provide a better match to the overall period ratio distribution \citep{2019arXiv190208772I}.
 
 \item On the other hand, the simulated multi-planet chains in the low-viscosity environment very closely match observations. In particular, our low-viscosity simulations can reproduce the large number of planets in the 3:2 and 2:1 configurations. This result is also not affected by outer giant planets, which can destabilise the inner systems \citep{2023A&A...674A.178B}, indicating that low-viscosity migration might be the major driver of wider resonances than the 5:4 (or tighter) resonances, which should therefore be more likely to form in discs with high viscosities.
 
\end{itemize}

Clearly, a large variety of parameters influence the outcome of planet-formation simulations. Here we have identified that different migration rates (caused by different viscosities) have consequences on the chains that are formed in these simulations. As the low-viscosity simulations better
match the chains than the high-viscosity simulations, and as the high-viscosity simulations better match the overall observed period ratios than the low-viscosity simulations, we suggest that the majority of the resonance chains are most probably formed at low viscosity, most likely without nearby external perturbers, as these tend to destroy the resonance chains effectively. In contrast, the majority of the systems in non-resonant configurations are likely to have formed in high-viscosity environments with a small number of external perturbers that allow a breaking of the chains rather than their total destruction. This interpretation indicates that a variety of viscosities (and outer companions) is probably needed to explain the different observed systems. Furthermore, this result is independent of the exact formation mechanism of the planets (e.g. pebbles or planetesimals), and of the initial embryo configuration, as long as the embryos can migrate and form resonance chains at the inner edge of the disc.

\begin{acknowledgements}

We thank the referee for their comments that helped us to improve our manuscript.

\end{acknowledgements}

\bibliographystyle{aa}
\bibliography{Stellar}

\newpage

\appendix
\section{Migration of single planets}
\label{app:migration}

\begin{figure*}
 \centering
 \includegraphics[scale=0.7]{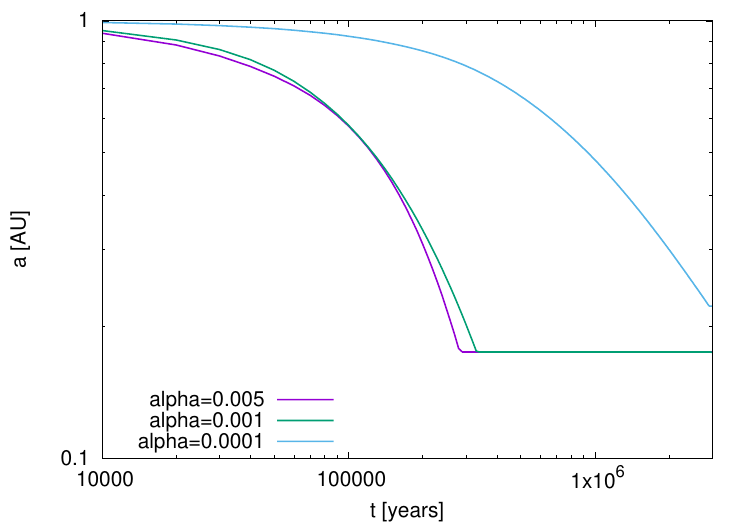}
 \includegraphics[scale=0.7]{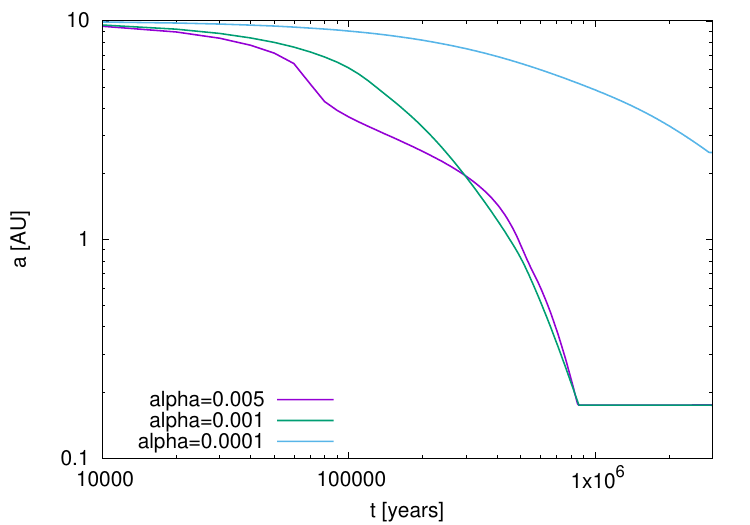} 
 \caption{Single migrating planets starting at 1 AU (left) or 10 AU (right) in discs with different viscosities. The viscosity of $\alpha=10^{-4}$ corresponds to our nominal simulations, while the viscosity of $\alpha=5\times 10^{-3}$ corresponds to the viscosity used in \citet{2019arXiv190208772I}.
   \label{fig:migration}
   }
\end{figure*}

We show in Fig.~\ref{fig:migration} the semi-major axis evolution of 10 Earth mass planets as a function of time. The planets start at either 1 or 10 AU and are embedded in discs with different viscosities. The planets at low viscosity migrate the slowest, while planets at higher viscosity migrate fastest, due to partial gap opening at low viscosity. Consequently, we expect the planets in the high-viscosity regime to result in tighter configurations due to the larger mutual migration velocities (see Appendix~\ref{app:2planets}).

\section{Migration rates}
\label{app:rates}

\begin{figure*}
 \centering
 \includegraphics[scale=0.7]{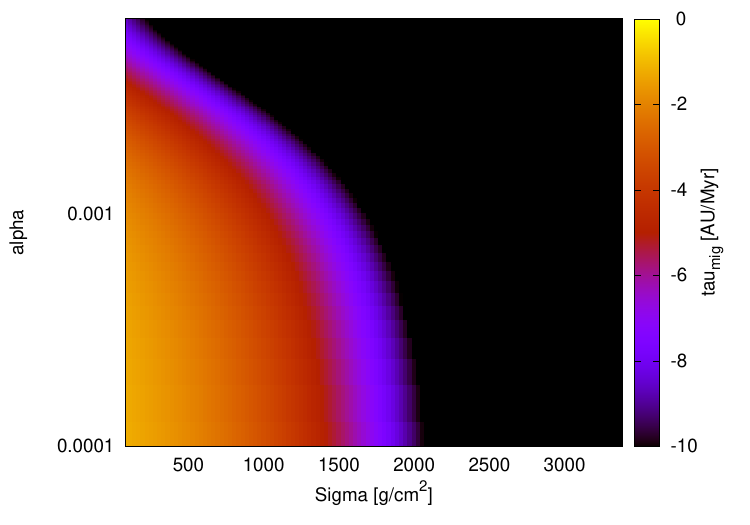}
 \includegraphics[scale=0.7]{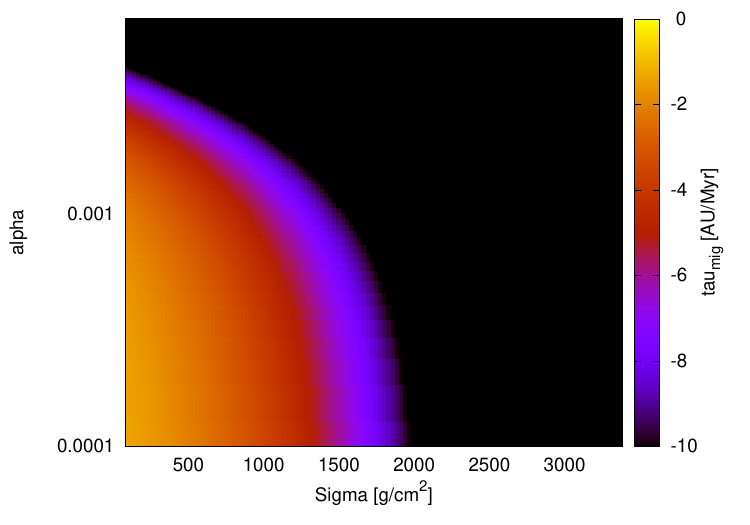} 
 \caption{Migration rates in AU per Myr of a 5 Earth mass planets (left) and 10 Earth mass planets (right) at 1 AU in a protoplanetary disc following the migration rate formula of \citet{2011MNRAS.410..293P} including a reduction of the migration rate by gap opening via the formalism of \citet{2018arXiv180511101K}.
   \label{fig:migmap}
   }
\end{figure*}

We show in Fig.~\ref{fig:migmap} the migration rate of 5 and 10 Earth mass planets in a ficticious disc model with

\begin{equation}
 \Sigma_{\rm g} = \Sigma_0 \left(\frac{r}{\rm AU}\right)^{-3/2}  \quad , \quad T = 170 \left(\frac{r}{\rm AU}\right)^{-1/2} {\rm K} \ ,
\end{equation}
where $\Sigma_0$ varies from 100 to 3400 $\frac{\rm g}{\rm cm^2}$. We additionally vary $\alpha$ from $10^{-4}$ to $8 \times 10^{-3}$. The migration rates are influenced by the gas surface density as well as by the viscosity.

As the gas surface density decreases, the migration rate decreases, because the torque acting on the planet is proportional to the gas surface density. A reduction by a factor of 3 in gas surface density therefore reduces the migration speed of the planets by a factor of 3 as well. 

In contrast, as the viscosity decreases, the planet can start to open a gap in the protoplanetary disc, which influences the migration velocity of the planet, where we follow the approach by \citet{2018arXiv180511101K}. \citet{2018arXiv180511101K} relate the type-II migration time-scale to the type-I migration time-scale (which we calculate as explained above) in the following way
\begin{equation}
\label{eq:migII}
 \tau_{\rm mig II} = \frac{\Sigma_{\rm up}}{\Sigma_{\rm min}} \tau_{\rm mig I} \ ,
\end{equation}
where $\Sigma_{\rm up}$ corresponds to the unperturbed gas surface density and $\Sigma_{\rm min}$ to the minimal gas surface density at the bottom of the gap generated by the planet. The ratio $\Sigma_{\rm up}/\Sigma_{\rm min}$ can be expressed through
\begin{equation}
\label{eq:Kgapopen}
 \frac{\Sigma_{\rm up}}{\Sigma_{\rm min}} = 1 + 0.04 K_{\rm mig} \ ,
\end{equation}
where
\begin{equation}
 K_{\rm mig} = \left( \frac{M_{\rm P}}{{\rm M}_\odot} \right)^2 \left( \frac{H}{r} \right)^{-5} \alpha^{-1} \ .
\end{equation}
Clearly, a smaller $\alpha$ values allows a faster opening of a gap and thus a reduction in the migration speed, as long as the gas surface density remains constant.

Low gas surface densities reduce the migration speeds and prevent the build up of chains, which is commonly assumed in the in-situ formation scenarios (e.g \citealt{2013ApJ...775...53H, 2015MNRAS.453.1471D, 2016ApJ...817...90L}). In these scenarios, the planets are normally fully formed before they start to migrate in the low gas surface density environments towards the end of the disc's lifetime. However, if the discs were very low in surface density at all time, it would be difficult to form planets in the first place due to the lack of planetary building blocks. If the disc had larger gas surface densities initially, the planets would form efficiently, but would then also start to migrate efficiently, resulting in the formation of planetary chains as in \citet{2019arXiv190208772I}. 

On the other hand, a reduction in viscosity avoids this issue. In fact a lower viscosity is even benefitial for the growth of planets due to the lower pebble scale height (e.g. \citealt{2014A&A...572A.107L, 2018A&A...609C...2B}). In the viscous accretion scenario, a lower viscosity would result in lower accretion rates of the disc, in potential conflict with observations (e.g. \citealt{2018MNRAS.474...88H, 2024A&A...689A.285S}). On the other hand, the accretion rates of discs could be driven by disc winds (e.g. \citealt{2023ASPC..534..465L}), which still allow a low mid-plane viscosity while maintaining large disc accretion rates. 

We note that the opening of a gap also depends on the disc's aspect ratio, where lower aspect ratios allow an easier opening of a gap. While the disc's aspect ratio scales with the power of -5, it's value normally does not change by large quantities. In our used disc model \citep{2015A&A...575A..28B}, the aspect ratio changes by about a factor of 2 within the first Myr of disc evolution. This would then amount to a similar change of $K_{\rm mig}$ as reducing the viscosity by a factor of 50. However, we implant the embryos into a disc that has already evolved for 2 Myr. At this stage the changes of the disc's aspect ratio (especially in the inner disc regions) are minimal. This is caused by the fact that the temperature of the disc is determined at these late stages by stellar heating, which also only changes minimally during these stages \citep{2015A&A...577A..42B}, resulting in a nearly constant aspect ratio in our model during planet formation \citep{2015A&A...575A..28B}. We thus think that evolution of migrating planets is mainly influenced by changes in the viscosity rather than the disc's surface or temperature.

\section{Low viscosity systems}
\label{app:systems}

\begin{figure*}
 \centering
 \includegraphics[scale=0.75]{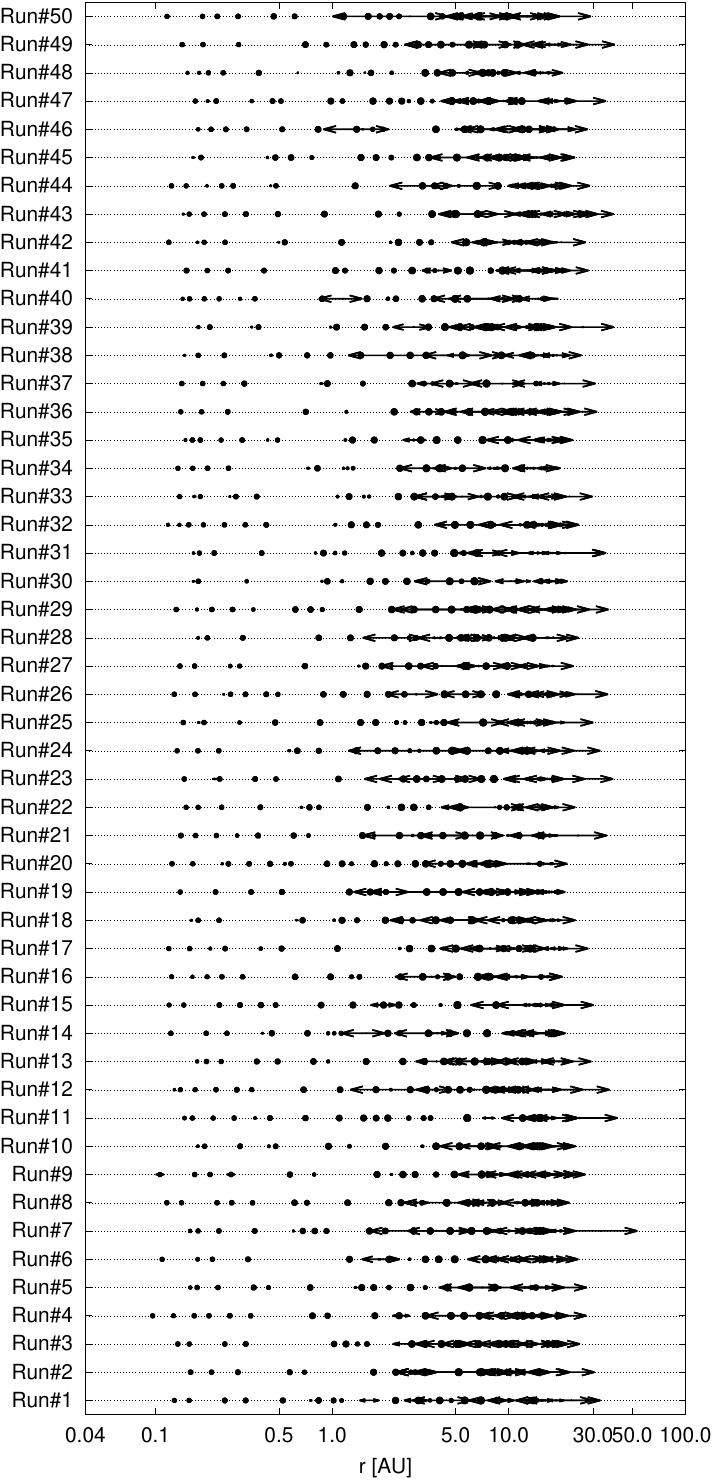}
 \includegraphics[scale=0.75]{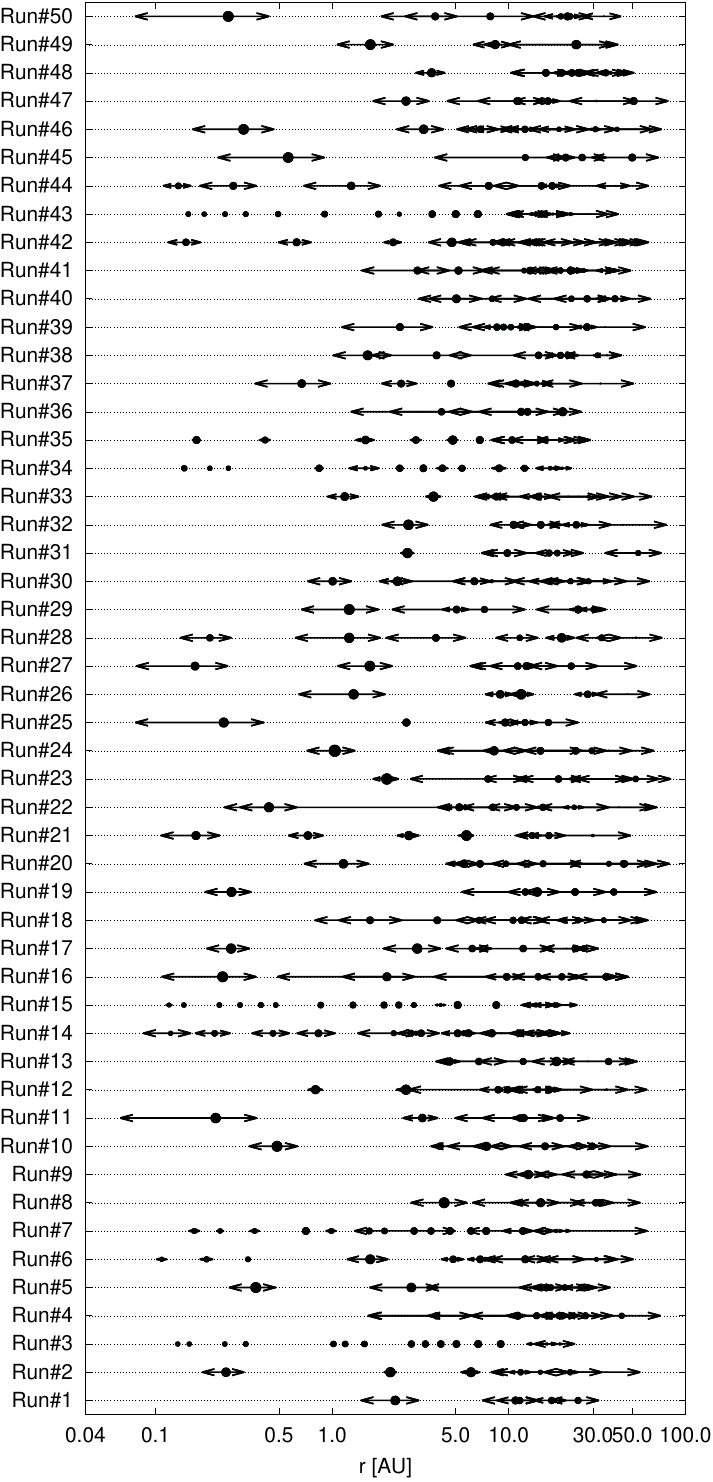} 
 \caption{Systems formed in our low-viscosity simulations after the gas-disc phase (left) and after 100 Myr of integration (right). The size of the dots is proportional to the planetary mass, while the arrow marks the aphelion and perihelion of the corresponding planet. We note that the synthetic observations only take planets up until 0.7 AU into account.
   \label{fig:systems}
   }
\end{figure*}

We show in Fig.~\ref{fig:systems} the configurations of the planetary systems after 3 Myr (end of the gas-disc phase) as well as after 100 Myr. The planetary systems become unstable after the gas-disc phase, which reduces the number of planets. We note that for the analysis presented in the main paper only planets more massive than 1 Earth mass are taking into account, because smaller planets are not reliably detectable.

\section{Influence of embryo number and disc viscosity}
\label{app:number}

The total number of planetary embryos can - of course - influence the outcome of the simulations, especially when looking at the limiting case of just a few planetary embryos, which might even be placed too far apart to gravitationally influence each other. While it is beyond the scope of this work to run an extensive suite of simulations with different number of initial embryos, we relate back to our previous work \citep{2023A&A...674A.178B}. In this work, we studied the formation of inner super-Earth systems with outer giant planets in the same set-up as presented here - with the exception that gas accretion onto planetary embryos was allowed so that gas giants can form.

\begin{figure}
 \centering
 \includegraphics[scale=0.7]{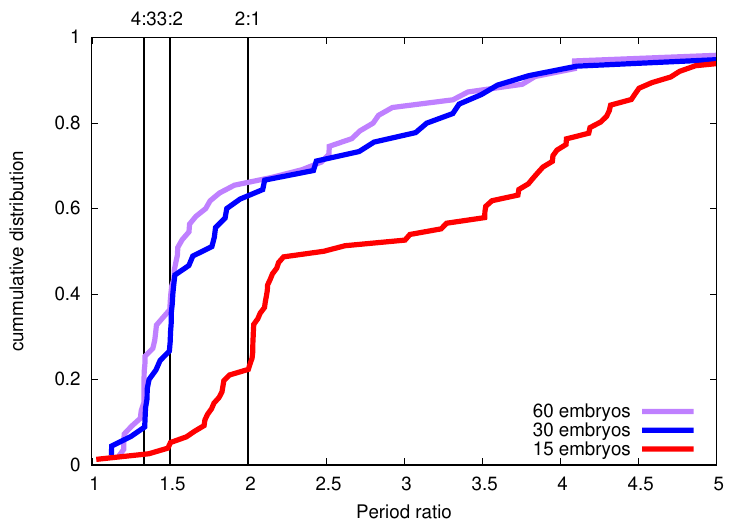}
 \caption{Period ratios of the planets interior to 1.0 AU in the simulations of \citet{2023A&A...674A.178B} with $K=5$ at the end of the gas-disc phase at 3 Myr. We note that the planets in the 15 embryo simulations mostly turn to gas giants, reducing their migration speeds and thus allow wider period ratios.
   \label{fig:periods}
   }
\end{figure}

In Fig.~\ref{fig:periods} we plot the period ratios of the systems within 1.0 AU for all planets in the simulations after the end of the gas-disc phase at 3 Myrs with initially 15, 30 or 60 planetary embryos and with a $K=5$ damping for the giant planets. We use here the full sample of planets, because some planets in the simulations of \citet{2023A&A...674A.178B} grew envelopes and became larger than 20 Earth masses - the limit used in the simulations by \citet{2019arXiv190208772I} as a definition of a super-Earth. We are here primarily interested in the influence of the number of embryos on the period ratios and not on their mass distribution.

Using 30 initial embryos results in a similar period ratio distribution compared to the results presented in Fig.~\ref{fig:beforeInstability}, namely a large fraction of systems in the 4:3 and 3:2 configuration. Using 60 initial planetary embryos results in a similar pile-up in the 4:3 and 3:2 configurations. This is in contrast to the simulations by \citet{2019arXiv190208772I}, which show a pile-up in tighter configurations. As \citet{2019arXiv190208772I} used 60 initial planetary embryos with the same starting configuration as our simulations, we can conclude that the viscosity is indeed responsible for the differences in the period ratios rather than the number of initial embryos. 

On the other hand, if only 15 initial embryos are used, the situation is different. Here we see a pile-up mostly in the 2:1 resonance configuration, rather than the 3:2 and 4:3. This is caused by the wide separation of the planetary embryos, where the embryos are spaced initially around 1.0 AU apart \citep{2023A&A...674A.178B}. Consequently only 1-2 planets reach the inner edge of the disc, where the outer planet already transitioned into a gas giant, which reduced their migration speed significantly so that the 2:1 configuration cannot be overcome. Furthermore, the large initial separation prevents the second planet from reaching the inner edge such that in most cases no resonant system is formed.

\begin{figure}
 \centering
 \includegraphics[scale=0.7]{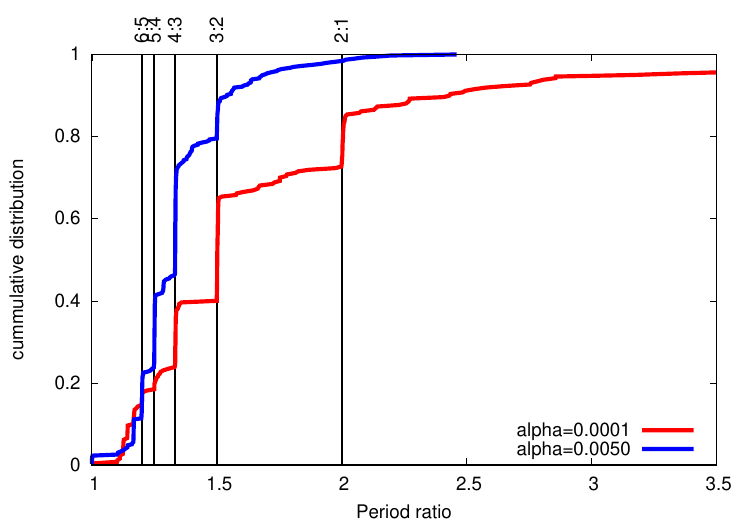}
 \caption{Period ratios of the planets interior to 1.0 AU for our simulations using $\alpha=10^{-4}$ and a comparison sample with the same disc properties, but $\alpha=5\times 10^{-3}$. The resulting periods are plotted after the end of the gas disc phase at 3 Myr.
   \label{fig:periodslowhigh}
   }
\end{figure}

In Fig.~\ref{fig:periodslowhigh} we show the period ratios of our simulations with $\alpha=10^{-4}$, as discussed in the main part of the paper, as well as from comparison simulations using the same disc and planet parameters, but $\alpha=5\times 10^{-3}$. Both sets of simulations feature 30 initial embryos. We show the periods at the end of the gas disc lifetime of 3 Myr. Clearly, the planetary systems formed in the high-viscosity simulations result in tighter resonant chains compared to the systems formed in the low-viscosity environments. In particular, the high-viscosity simulations feature a larger fraction of planets in the 4:3 and 5:4 (and tighter) resonance configurations, while the low-viscosity simulations produce many more planets in the 3:2 and 2:1 resonance configurations. We note that the exact period configurations are subject to change after the instability phase, but instabilities break the resonances and do not populate the resonance configurations, indicating that overall trend that higher viscosity systems produce resonant chains in tighter configurations compared to low-viscosity simulations will continue to hold.

We can conclude from these results that as long as enough planetary embryos are present,  high viscosities are needed to make tight chains, while lower viscosities are needed to make wider chains (e.g. for the 3:2 rather than 5:4 configuration). However, if the number of planetary embryos is very limited, even wider configurations are possible. Here, we have to admit the caveat that this might be caused by the larger planetary masses for the simulations with 15 initial embryos that allow deeper gap opening in the simulations of \citet{2023A&A...674A.178B} compared to our here presented simulations.

\section{Planetary masses}
\label{app:masses}

\begin{figure}
 \centering
 \includegraphics[scale=0.53]{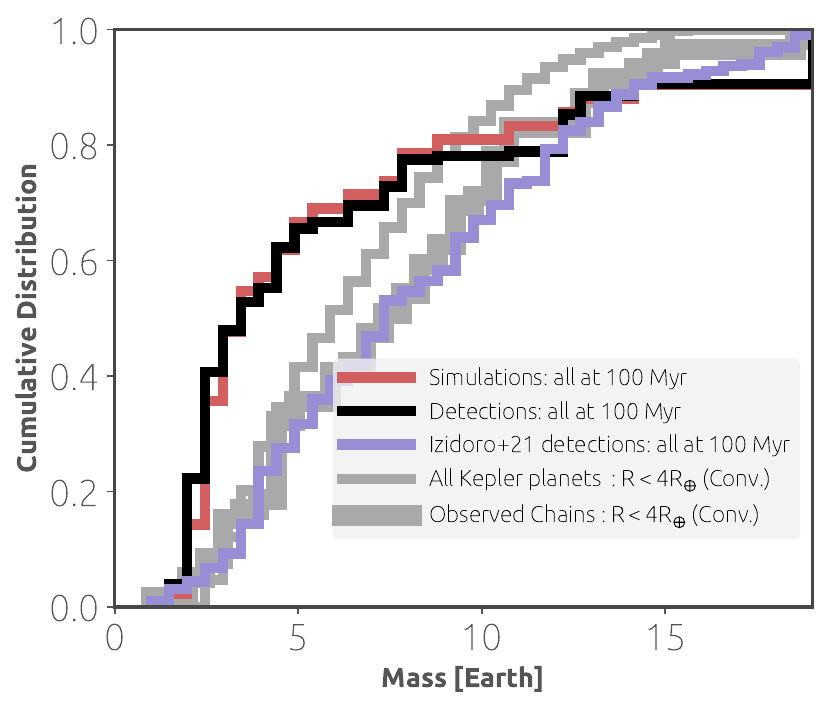}
 \includegraphics[scale=0.7]{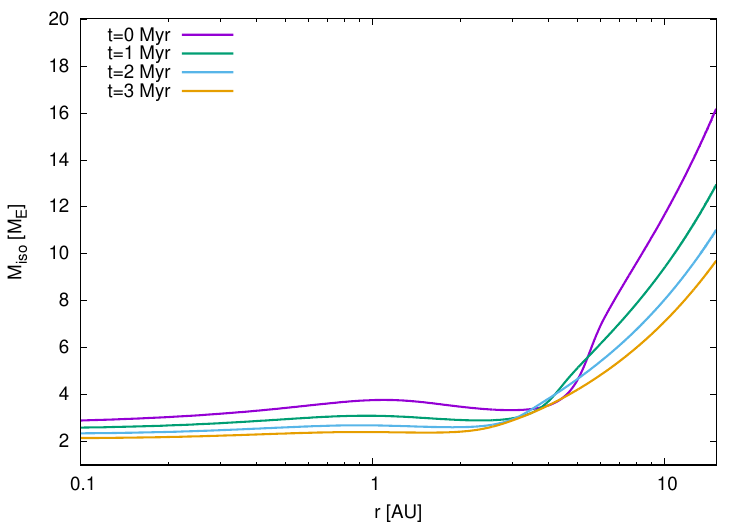} 
 \caption{Top: Cumulative distribution of the planetary masses in our simulations (red) and their detections (black) as well as from the simulations of \citet{2019arXiv190208772I} (purple) after 100 Myr of integration time. We also overplot the masses from the \textit{Kepler} observations using the mass-radius relationship of \citet{2016ApJ...825...19W}. Bottom: Time evolution of the pebble isolation mass in our model, following the recipe of \citet{2018arXiv180102341B}. Here $t=0$ corresponds to the starting time of the simulations, so to a disc that is already 2 Myr old in the framework of \citet{2015A&A...575A..28B}.
   \label{fig:masses}
   }
\end{figure}

The planetary mass determines not only the migration speed, but are also important for the stability of planetary systems (e.g. \citealt{1993Icar..106..247G, 1996Icar..119..261C}). In Fig.~\ref{fig:masses} we show the cumulative mass distribution of our planets after 100 Myr of evolution. Clearly, the masses of our low-viscosity planetary systems are too low compared to the inferred masses from the \textit{Kepler} observations (using the mass-radius relationship of \citet{2016ApJ...825...19W}). The high-viscosity simulation of \citet{2019arXiv190208772I}, on the other hand, match very well with the observations, especially for the masses of the chains. 

In our model, the only difference between the high viscosity and low-viscosity simulations is the migration velocity. The viscosity enters also into the pebble isolation mass \citep{2018arXiv180102341B}, but is set to the same value in all our simulations to allow an easier comparison. Thus, the difference in mass only arises from the difference in migration velocity. As shown in Fig.~\ref{fig:migration}, the planets formed in low-viscosity environment migrate slower and can not reach the inner edge of the disc towards the end of the disc's lifetime, if they start to migrate from exterior to $\approx 6-7$ AU (see also \citealt{2017AJ....153..222B}). However, in this region, the pebble isolation mass starts to increase dramatically (see right in Fig.~\ref{fig:masses}). Therefore the majority of the planets that form via pebble accretion in the low-viscosity environments that migrate to the inner disc only reach up to $\approx$3 Earth masses. As pebble accretion is very efficient, nearly all planets reach this mass, explaining the lower mass tail in the left of Fig.~\ref{fig:masses}. The larger planetary masses are caused by collisions after the gas disc phase during the instability phase of the systems.

In contrast, the planets formed in simulations with higher viscosity can migrate from much further out into the inner disc region due to their faster migration velocity (Fig.~\ref{fig:migration}). This result in a steady influx of planets that have reached higher masses due to the higher pebble isolation mass exterior to a 6-7 AU. Furthermore, these planets are massive enough that they can collide already during the gas disc-phase \citep{2019arXiv190208772I}, which is not the norm for the planets formed in the low-viscosity simulations. Consequently, the planets formed in the high-viscosity simulations in the inner disc feature higher planetary masses, even though the formula for the pebble isolation mass is the same.

\section{Period ratios}
\label{app:periods}

\begin{figure}
 \centering
 \includegraphics[scale=0.53]{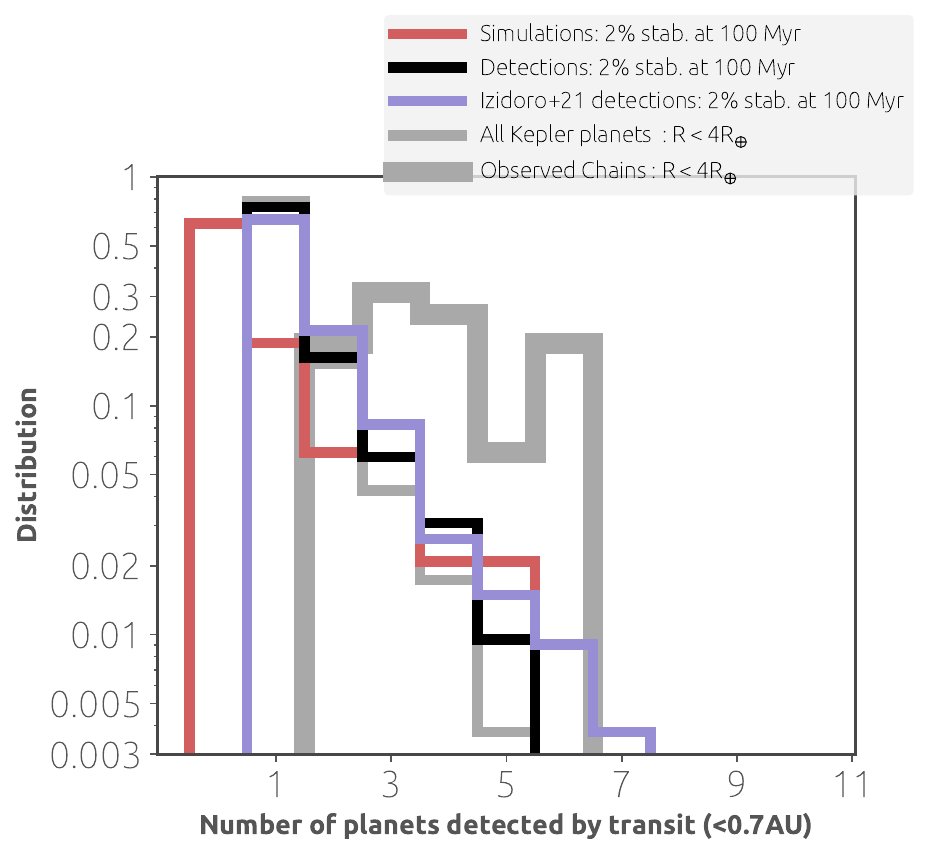}
 \includegraphics[scale=0.53]{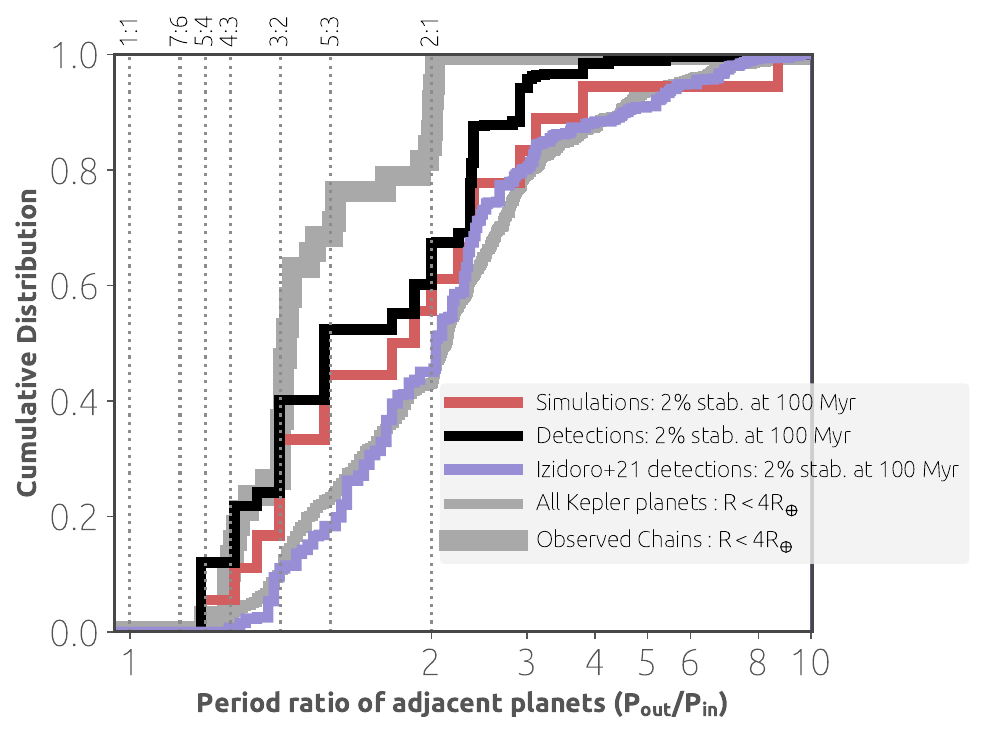} 
 \caption{Number of synthetically observed planets from our simulations (top) and the period ratios of adjacent planets (bottom) for a mix of planetary systems that includes 2\% stable chains and 98\% systems that underwent instabilities. The colors are the same as in Fig.~\ref{fig:beforeInstability}. Both sets of simulations give a nice match to the number of detected planets, but the period ratios are matched better in the high-viscosity case, compared to the low-viscosity case.
   \label{fig:mixsafter}
   }
\end{figure}

We show in Fig.~\ref{fig:mixsafter} a mixture of 2\% stable chains and 98\% systems that underwent instabilities. This mixture was used in the past high-viscosity simulations of \citet{2019arXiv190208772I}, as it gave the best fit to the observations (indicated by the purple line). The low-viscosity simulations still fail to reproduce the observed period ratios, as their period ratios remain too tight.

\section{Resonant chains of two-planet systems}
\label{app:2planets}

\begin{figure}
 \centering
 \includegraphics[scale=0.7]{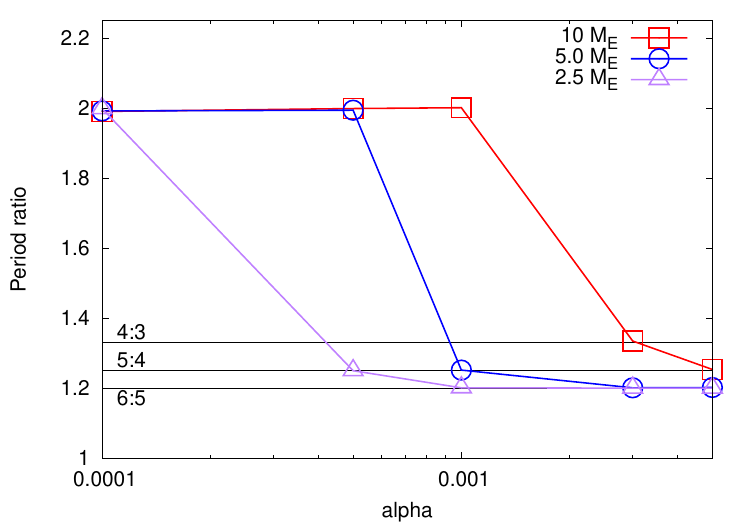}
 \caption{Final period ratios of two-planet systems with equal masses in discs with different viscosities.
   \label{fig:2planets}
   }
\end{figure}

In order to investigate the influence of viscosity on the period ratio of planetary systems, we conduct a simple experiment. We take systems with 2 equal mass planets and let them migrate through a discs, where only the viscosity varies between the different set-ups. Different viscosities can cause different migration velocities (see Appendix~\ref{app:migration}) due to the change of the strength of the entropy driven corotation torque (e.g. \citealt{2011MNRAS.410..293P, 2013A&A...550A..52B}). In particular, high-viscosity environments can lead to outward migration, if the disc's surface density and temperature gradients are in the right regime. In fact, high viscosities will lead to outward migration in our used disc model \citep{2015A&A...575A..28B}, which complicates our simple experiment. We thus use a simple power law disc model for this experiment:
\begin{equation}
 \Sigma_{\rm g} = 1700 \left(\frac{r}{\rm AU}\right)^{-3/2} \frac{\rm g}{\rm cm^2} \quad , \quad T = 170 \left(\frac{r}{\rm AU}\right)^{-1/2} {\rm K} \ .
\end{equation}
This simple disc model does not support outward migration for all levels of viscosity, allowing us to study clearly the migration behaviour of pairs of planets of equal mass, which are not growing. We start our planetary systems just outside the 2:1 resonance in order to study how the different migration velocities influence the final resonance configuration. We use here the same migration prescription as in our main paper \citep{2011MNRAS.410..293P}.

We show in Fig.~\ref{fig:2planets} the final period ratios of our planetary systems as function of the disc's viscosity. An increase in the $\alpha$-viscosity parameter results in tighter resonance configurations, caused by the larger migration velocities of the planets in the higher viscosity environments. Furthermore, lower planetary masses result in tighter resonances as well, compared to higher mass planets. The reason for this is the strength of the resonance, which scales with the planetary masses, and needs to be overcome by migration.

The here observed trend is reflected in our main simulations, where the simulations at low viscosity with low masses show wider final period ratios compared to the simulations with higher masses and higher viscosities. We note though that the exact period ratio in a multi-planet system is more complicated as in the two-planet case, due to the increase in eccentricities caused by the gravitational interaction of multiple planets that can influence the exact resonance configuration.

\end{document}